\documentclass[fleqn,usenatbib]{mnras}

\usepackage{newtxtext,newtxmath}

\usepackage[T1]{fontenc}

\DeclareRobustCommand{\VAN}[3]{#2}
\let\VANthebibliography\thebibliography
\def\thebibliography{\DeclareRobustCommand{\VAN}[3]{##3}\VANthebibliography}

\usepackage{siunitx,subfig,xspace,natbib}
\usepackage{hyperref}
\usepackage{amsfonts}%
\usepackage{mathrsfs}
\usepackage{bbm}
\usepackage{cprotect}

\usepackage{graphicx}	
\usepackage{amsmath}	
\usepackage{amssymb}	

\newcommand{\mrm}[1]{\mathrm{#1}}
\newcommand{\nuc}[2]{$\mrm{^{#2}#1}$}

\title[Dark Matter in Reticulum II]{Reticulum II: Particle Dark Matter and Primordial Black Holes Limits}

\author[Thomas Siegert et al.]{
Thomas Siegert,$^{1}$\thanks{E-mail: tho.siegert@gmail.com}
Celine Boehm,$^{2}$
Francesca Calore,$^{3}$
Roland Diehl,$^{4}$
Martin G. H. Krause,$^{5}$
\newauthor
~Pasquale D. Serpico,$^{3}$
and Aaron C. Vincent$^{6}$
\\
$^{1}$Institut f\"ur Theoretische Physik und Astrophysik, Universit\"at W\"urzburg, Campus Hubland Nord, Emil-Fischer-Str. 31, 97074 W\"urzburg, Germany\\
$^{2}$Sydney Consortium for Particle Physics and Cosmology, School of Physics, The University of Sydney, NSW 2006, Australia\\
$^{3}$Univ. Grenoble Alpes, Univ. Savoie Mont Blanc, CNRS, LAPTh, F-74940 Annecy, France\\
$^{4}$Max-Planck-Institut f\"ur extraterrestrische Physik, Gie\ss enbachstra\ss e, 85741 Garching b. M\"unchen, Germany\\
$^{5}$Centre for Astrophysics Research, Department of Physics, Astronomy and Mathematics, University of Hertfordshire, College Lane, Hatfield, Hertfordshire AL10 9AB, UK\\
$^{6}$Department of Physics, Enigneering Physics and Astronomy, Queen’s University, Kingston, ON, K7L 3N6, Canada
}

\date{Accepted XXX. Received \today; in original form ZZZ}

\pubyear{2021}

\begin{document}
\label{firstpage}
\pagerange{\pageref{firstpage}--\pageref{lastpage}}
\maketitle

\begin{abstract}
Reticulum II (Ret\,II) is a satellite galaxy of the Milky Way and presents a prime target to investigate the nature of dark matter (DM) because of its high mass-to-light ratio.
We evaluate a dedicated INTEGRAL observation campaign data set to obtain $\gamma$-ray fluxes from Ret\,II and compare those with expectations from DM.
Ret\,II is not detected in the $\gamma$-ray band 25--8000\,keV, and we derive a flux limit of $\lesssim 10^{-8}\,\mathrm{erg\,cm^{-2}\,s^{-1}}$.
The previously reported 511\,keV line is not seen, and we find a flux limit of $\lesssim 1.7 \times 10^{-4}\,\mathrm{ph\,cm^{-2}\,s^{-1}}$.
We construct spectral models for primordial black hole (PBH) evaporation and annihilation/decay of particle DM, and subsequent annihilation of $e^+$s produced in these processes.
We exclude that the totality of DM in Ret\,II is made of a monochromatic distribution of PBHs of masses $\lesssim 8 \times 10^{15}\,\mathrm{g}$.
Our limits on the velocity-averaged DM annihilation cross section into $e^+e^-$ are $\langle \sigma v \rangle \lesssim 5 \times 10^{-28} \left(m_{\rm DM} / \mathrm{MeV} \right)^{2.5}\,\mathrm{cm^3\,s^{-1}}$.
We conclude that analysing isolated targets in the MeV $\gamma$-ray band can set strong bounds on DM properties without multi-year data sets of the entire Milky Way, and encourage follow-up observations of Ret\,II and other dwarf galaxies.
\end{abstract}




\section{Introduction}\label{sec:intro}
\citet{Siegert2016_dsph} reported a $3.1\sigma$ signal of a 511\,keV line from the direction of the dwarf galaxy Reticulum II \citep[Ret\,II; distance $d = 30 \pm 2$\,kpc;][]{Koposov2015_dsph}.
The line flux from this previous work was extraordinarily high, $F_{511} = (1.7 \pm 0.5) \times 10^{-3}\,\mathrm{ph\,cm^{-2}\,s^{-1}}$, which would make Ret\,II the most luminous 511\,keV source in the sky.
The signal has been interpreted in terms of a recent binary neutron star (NS) merger \citep{Fuller2018_511NS}, possibly outshining the entire galaxy.
Ret\,II has also been reported to show an excess at GeV energies \citep{Geringer-Sameth2015_RetII} which might point to a dark matter (DM) origin of the signal.
Follow-up analyses found no flux enhancement above the diffuse GeV emission \citep{Albert2017_Fermi_dsph_DM}.
Intrigued by these finding, the INTEGRAL \citep{Winkler2003_INTEGRAL} satellite performed a dedicated observation campaign to follow up the signal.
In this work, we analyse the new 1\,Ms data together with archival data from the coded-mask $\gamma$-ray spectrometer onboard INTEGRAL, SPI \citep{Vedrenne2003_SPI}.
We use our findings to set limits on possible DM models, including primordial black holes (PBHs) and annihilation or decay of (light) DM particles.

This paper is structured as follows:
In Sec.\,\ref{sec:new_analysis}, we describe the new observations, and how SPI data are analysed in general.
We build photon-emission models expected from the evaporation of PBHs and annihilating/decaying particle DM in Sec.\,\ref{sec:models}.
Here, we also include the possible annihilation of positrons ($e^+$s) as a DM product into the galaxy emission spectra.
Our results are presented in Sec.\,\ref{sec:results}.
We conclude in Sec.\,\ref{sec:conclusion}.

\section{New Observation Campaign}\label{sec:new_analysis}
The previous analysis by \citet{Siegert2016_dsph} focussed on the search of 511\,keV signals from the 39 satellite galaxies of the Milky Way (MW) known at the time.
In their data set, the entire sky was modelled and the diffuse 511\,keV emission from the MW itself was included.
Before the new observation campaign, Ret\,II had never been targeted for observations by SPI because 1) the galaxy was only discovered in 2015 \citep{Koposov2015_dsph}, and 2) other known X- or $\gamma$-ray sources are separated from Ret\,II by at least $18^{\circ}$.
However, thanks to SPI's wide fully-coded field of view (FCFOV) of $16^{\circ} \times 16^{\circ}$ ($30^{\circ} \times 30^{\circ}$ partially coded), Ret\,II data were collected also before its discovery.
Outside the full coding region, the effective area of the instrument sharply drops which placed previous Ret\,II observations right onto exposure edges with a cumulative on-target observation time of $550$\,ks, split over ten years.
While these effect were taken into account in the analysis, dedicated observations, pointed directly to Ret\,II, would be suitable to validate earlier findings.

\subsection{Data Set}\label{sec:data_set}
Our new data set includes previous observations between 2003 and 2013 and adds $\sim 1$\,Ms during the Ret\,II campaign in 2018.
We selected all pointed observations (pointings) whose distance to Ret\,II is less than $16^{\circ}$ so that our target is always in the FCFOV.
This amounts to 603 pointings with an average observation time of 3300\,s each.
The mean dead time of a working detector is about 18.2\,\%.
Since the launch of the INTEGRAL mission in October 2002, four of the 19 SPI detectors have failed, reducing the effective area by $\approx 20\,\%$.
The total dead-time- and efficiency-corrected exposure time then amounts to 1.38\,Ms.
We use SPI's bandpass between 25 and 8000\,keV, and perform our analysis in 29 logarithmic energy bins, plus a single narrow bin for the 511\,keV line from 508 to 514\,keV.

\subsection{General Analysis Method}\label{sec:spi_analysis}
SPI data analysis relies on the comparison of expected detector illuminations, that is, patterns in the 19-detector data space, with the measured count rate per pointing and energy.
Fluxes are estimated by maximising the Poisson likelihood,
\begin{equation}
	\mathscr{L}(D|M) = \prod_{p} \frac{m_p^{d_p} \exp(-m_p)}{d_p!}\mathrm{,}
	\label{eq:likelihood}
\end{equation}
\noindent where $d_p$ are the measured counts in pointing $p$, and $m_p$ is the model expectation.
MeV $\gamma$-ray measurements are described by a two component model, one representing the instrumental background $m^{\rm BG}$, and one representing the celestial emission $m^{\rm SKY}$.
Instrumental background originates from cosmic-ray interactions with the satellite material and leads to continuum ($C$) and line ($L$) backgrounds \citep{Diehl2018_BGRDB}.
Spatial emission models are convolved with the imaging response function of SPI, that is, the mask coding, leading to a 19-element vector of expected source counts for each pointing.
The total model reads
\begin{eqnarray}
	m_{p} & = & m_{p}^{\rm BG} + m_{p}^{\rm SKY} = \nonumber\\
	& = & \sum_{t,t'} \left[ \sum_{i \in \left\{C,L\right\}} \beta_{i,t} R_{p,i}^{\rm BG} + \sum_{j=1}^{N_S} \alpha_{j,t'} R_{p;lb}^{\rm SKY} M_j^{lb}(\boldsymbol{\phi_j}) \right]\mathrm{,}
	\label{eq:spimodfit_model}
\end{eqnarray}
\noindent where $R_{p,i}^{\rm BG}$ is the background model (response) appropriate for pointing $p$, scaled by the amplitudes $\beta_{i,t}$ for background components $i \in \left\{C,L\right\}$ varying on timescales $t$, $R_{p;lb}^{\rm SKY}$ is the coded-mask response in pointing $p$, converting the $j = 1 \dots N_S$ sky models $M_j^{lb}(\boldsymbol{\phi})$ from image space $lb$ to the data space.

The source models may have additional parameters $\boldsymbol{\phi_j}$, for example defining the position of point sources, and are scaled by the amplitudes $\alpha_{j,t'}$, possibly varying on another time scale $t'$.
Given SPI's angular resolution of $2.7^{\circ}$, and the expected maximum extent of Ret\,II's DM halo of $1^{\circ}$, we model the galaxy as point source,
\begin{equation}
	M_{\rm Ret\,II}(l,b) = \delta(l-l_0)\delta(b-b_0)\mathrm{,}
	\label{eq:point_source}
\end{equation}
\noindent with the Galactic coordinates of Ret\,II $(l_0,b_0) = (266.30, -49.74)$.
All other sources that can be expected in our data set, EXO\,0748-676, ESO\,33-2, SMC\,X-1, and LMC\,X-4 are modelled as point sources as well.
The fitted parameters in Eq.\,(\ref{eq:spimodfit_model}) are $\beta_{i,t}$ and $\alpha_{j,t'}$.
The fit is performed with \verb|spimodfit| \citep{spimodfit}, which creates a spectrum by looping over the energy bins defined in our data set.
Details about the procedure are found in \citet{Siegert2019_SPIBG}.

\subsection{Spectral Fitting}\label{sec:spectral_fits}
The spectrum created in this way is subject to the instrument dispersion, that is, the probability that a photon with initial energy $E_i$ is measured at a final energy $E_f$ owing to scattering in the instrument.
In order to correct for the intrinsic dispersion, expected spectral models $\frac{dF}{dE}$ are convolved with the energy-redistribution matrix $H(E_i,E_f)$ that forward-folds the models into the appropriate data space.
Thus, the spectral models read
\begin{equation}
	\frac{dF\left(E_f; \boldsymbol \psi \right)}{dE_f} = \int\,dE_i H(E_i,E_f) \frac{dF\left(E_i; \boldsymbol \psi \right)}{dE_i}\mathrm{,}
	\label{eq:folded_model}
\end{equation}
\noindent with $\frac{dF}{dE_f}$ being the folded model which is compared to the extracted flux values from Sec.\,\ref{sec:spi_analysis}, and $\frac{dF}{dE_i}$ is the intrinsic source model that depends on a set of spectral parameters $\boldsymbol \psi$.
We use the Multi-Mission-Maximum-Likelihood \citep[3ML,][]{Vianello2015_3ML} framework to perform spectral fits.
Details about the choices of prior probabilities and parameter ranges are given in Appendix\,\ref{sec:appendix_spec_fits}.

We note that, ideally, the steps explained in Secs.\,\ref{sec:spi_analysis} and \ref{sec:spectral_fits} should be performed in one single step to enhance the sensitivity of the parameter estimation, which is, however, not yet implemented in the current software release.

\section{Expectations in the MeV Band}\label{sec:models}
No MeV emission has ever been detected from a (dwarve) satellite galaxy.
Therefore, the expected signals have a large variety, ranging from $\gamma$-ray line emission due to radioactive decays and (subsequent) $e^+$ emission.
Population synthesis models estimate a 1.8\,MeV $\gamma$-ray line flux from the decay of \nuc{Al}{26} of the order of $10^{-6}\,\mrm{ph\,cm^{-2}\,s^{-1}}$ from the Large Magellanic Cloud (LMC), about one order of magnitude below SPI's sensitivity \cite{Diehl2018_BGRDB}.
Because the expected line flux is related to the star formation and supernova rate, any other satellite galaxy, and in particular Ret\,II, can be expected to show no significant excess at 1.8\,MeV.
\citet{Cordier2004_511dm} performed a search for a 511\,keV line owing to dark matter annihilation/decay from the Sgr dwarve galaxy, but found no excess.
\citet{Siegert2016_dsph} extended this search to all satellite galaxies of the MW, which resulted in one $3\sigma$ signal from the direction of Ret\,II out of 39 tested galaxies.
Where this signal, if true, comes from is unknown, and might point to different origins in terms of NS mergers \citep{Fuller2018_511NS}, accreting X-ray binaries \citep{Siegert2016_V404}, or DM \citep{Boehm2004_dm}.
All these scenarios would show additional emission features in the MeV band besides a sole 511\,keV line.
In particular additional broad $\gamma$-ray lines from other nucleosynthesis products in NS mergers, continuum emission from a population of X-ray binaries, or radiative corrections in the final products from DM decay/annihilation or PBH decay.
Therefore, we extract the spectrum of Ret\,II and nearby sources using the logarithmic energy binning defined in Sec.\,\ref{sec:data_set}, plus a single bin for the 511\,keV line.
In this way, we can assess the previously found signal from Ret\,II in a model-independent fashion, and determine model dependent parameters using the additionally-expected components in Sec.\,\ref{sec:model_expectations}.

\subsection{Modelling the MeV Band in Ret\,II}\label{sec:model_expectations}
We focus on the $\gamma$-ray spectra expected from the evaporation of PBHs in the mass range $10^{14}$--$10^{18}$\,g and annihilating/decaying particle DM in the mass range $0.05$--$300\,\mathrm{MeV}$.
Other signals, for example the $\gamma$-ray emission from old NS mergers remnants, are not part of our predictions because the expected emission of one 10\,kyr old remnant in Ret\,II is about one million times lower \citep{Korobkin2020_NSmerger} than the instrument sensitivity.
While emission from GeV DM would also imprint in the MeV band, SPI is most sensitive to annihilation/decay to $e^+e^-$ and subsequent emission processes considering the final state radiation (FSR) of the pairs as well as the annihilation of $e^+$s during propagation and after thermalisation.
Therefore, we restrict our models to the two channels $e^+e^-$ and $\gamma\gamma$, and present one $\mu^+\mu^-$ example for comparison.
We follow the formalism of \citet{Fortin2009_DMGammaRaySpectra} throughout the paper.

The emission from DM haloes is described by a `particle physics' and an `astrophysics' component as
%

\begin{equation}
	\frac{dF_{\gamma}\left(E_{\gamma}; \kappa, M, \boldsymbol{\psi}, n \right)}{dE_{\gamma}} = \underbrace{\frac{\kappa}{4\pi M} \frac{dN_{\gamma}\left(E_{\gamma}; \boldsymbol{\psi}\right)}{dE_{\gamma}}}_{\mathrm{particle\,physics:\,PBH\,or\,DM}} \underbrace{\int_{\rm los}\,dr d\Omega \rho_{\rm halo}^n(r,\Omega)}_{\mathrm{astrophysics:\,D\,or\,J}}\mathrm{,}
	\label{eq:astroflux}
\end{equation}
\noindent where $\kappa$, $M$, and $\boldsymbol{\psi}$ describe the parameters of interest for the specific cases of PBH evaporation, self-conjugated DM annihilation or decay (see Tab.\,\ref{tab:DM_params_flux}), and the last term is the D- ($n=1$) or J-factor ($n=2$) of the galaxy.
They describe the extent of the emission, calculated by an integration of the galaxy's halo profile $\rho_{\rm halo}(r,\Omega)$ over the line of sight (los).
\begin{table}
	\centering
	\begin{tabular}{l|cccc}
		\hline\hline
		Model & $\kappa$ & $M$ & $n$ & $\boldsymbol{\psi}$ \\
		\hline
		PBH & $f_{\rm PBH}$ & $M_{\rm BH}$ & $1 \rightarrow D$ & $\left\{f_{\rm Ps}, E_{\rm kin}^{\rm max}\right\}$  \\
		DM Annihilation & $\langle \sigma v \rangle$ & $2m_{\rm DM}$ & $2 \rightarrow J$ & $\left\{f_{\rm Ps}, E_{\rm kin}^{\rm max}\right\}$ \\
		DM Decay & $1/\tau$ & $m_{\rm DM}$ & $1 \rightarrow D$ & $\left\{f_{\rm Ps}, E_{\rm kin}^{\rm max}\right\}$ \\	
		\hline
	\end{tabular}
	\caption{Parameters for considered DM models in Eq.\,(\ref{eq:astroflux}).}
	\label{tab:DM_params_flux}
\end{table}
The photon spectrum $\frac{dN_{\gamma}(E_{\gamma};\boldsymbol{\psi})}{dE_{\gamma}}$ is related to the photon density of states as $\frac{1}{N_{\gamma}}\frac{dN_{\gamma}}{dE_{\gamma}} = \frac{1}{\kappa}\frac{d\kappa}{dE_{\gamma}}$ so that the number of emitted photons per process is $N_{\gamma} = \int\,dE_{\gamma} \frac{dN_{\gamma}}{dE_{\gamma}}$.
$N_{\gamma}$ is thus normalising the source spectrum, Eq.\,(\ref{eq:astroflux}), in terms of the totally-emitted photons.

We use the range of D- and J-factors of Ret\,II from the literature \citep{Bonnivard2015_RetII,Evans2016_Jfactors,Albert2017_Fermi_dsph_DM}, in particular $D \approx (1$--$4) \times 10^{18}\,\mathrm{GeV\,cm^{-2}}$, and $J \approx (0.2$--$3.7) \times 10^{19}\,\mathrm{GeV^2\,cm^{-5}}$. The conversion of the D- and J-factor into other often-used units are $D = (0.9$--$3.5) \times 10^{-2}\,\mathrm{M_{\odot}\,pc^{-2}} = (1.9$--$7.3) \times 10^{-6}\,\mathrm{g\,cm^{-2}}$, and $J=(0.6$--$8.4) \times 10^{-3}\,\mathrm{M_{\odot}^2\,pc^{-5}} = (0.8$--$11.8) \times 10^{-29}\,\mathrm{g^2\,cm^{-5}}$.
We use these ranges as uniform priors in our spectral fits to include uncertainties in the DM halo as well as the distance to Ret\,II.
Depending on the DM model, the parameters of interest change and are listed in Tab.\,\ref{tab:DM_params_flux}.
Here, $f_{\rm PBH}$ is the fraction of DM made of PBHs (unitless), $\langle \sigma v \rangle$ is the velocity-averaged DM annihilation cross section in units of $\mathrm{cm^3\,s^{-1}}$, $\tau = \Gamma^{-1}$ is the decay time of a DM particle in units of $\mathrm{s}$, $M_{\rm BH}$ is the mass of PBHs in units of $\mathrm{g}$, and $m_{\rm DM}$ is the DM particle mass in units of $\mathrm{keV}$.
The additional spectral model parameters $\boldsymbol{\psi} = \left\{f_{\rm Ps}, E_{\rm kin}^{\rm max}\right\}$ appear when $e^+$ annihilation is included (see below).

\subsubsection{Primordial Black Hole Evaporation}\label{sec:PBH_evaporation}
The spectrum of an evaporating black hole (BH) of mass $M_{\rm BH}$ due to Hawking radiation \citep{Hawking1975_HawkingRadiation,Page1976_PBHHawking} is
\begin{equation}
	\left(\frac{dN_{i}}{dE_{i}}\right)_{\rm BH} = \frac{1}{2\pi} \frac{\Gamma_i(E_{i},M_{\rm BH})}{\exp\left(\frac{E_{i}}{T_{\rm BH}}\right) - (-1)^{2s_i}}\,\mathrm{.}
	\label{eq:hawking_emission}
\end{equation}
In Eq.\,(\ref{eq:hawking_emission}), $E_i$ is the energy of particle $i$ and $s_i$ its spin.
$\Gamma_i(E_i,M_{\rm BH})$ is the `greybody' factor that alters the expected blackbody distribution of possible emitted particles, given the BH temperature
\begin{equation}
	T_{\rm BH} = \frac{\hbar c^3}{8 \pi G k_B M_{\rm BH}} = 6 \times 10^{-8} \left(\frac{M_{\odot}}{M_{\rm BH}}\right)\,\mathrm{K} = 1.06 \left(\frac{10^{16}\,\mathrm{g}}{M_{\rm BH}}\right)\,\mathrm{MeV,}
	\label{eq:BH_temperature}
\end{equation}
\noindent where $\hbar$ is reduced Planck's constant, $G$ is the gravitational constant, $c$ is the speed of light, and $k_B$ is Boltzmann's constant.

We use \verb|BlackHawk| \citep{Arbey2019_BlackHawk} to calculate the spectra of all relevant particles in the energy range between 1\,keV and 1\,GeV.
\verb|BlackHawk| allows to include secondary particle production due to hadronization, fragmentation, decay, and other processes as a result of BH evaporation, which we add to our spectra.
\begin{figure}
	\centering
	\includegraphics[width=\columnwidth,trim=0.0cm 0.0cm 0.0cm 0.0cm, clip=true]{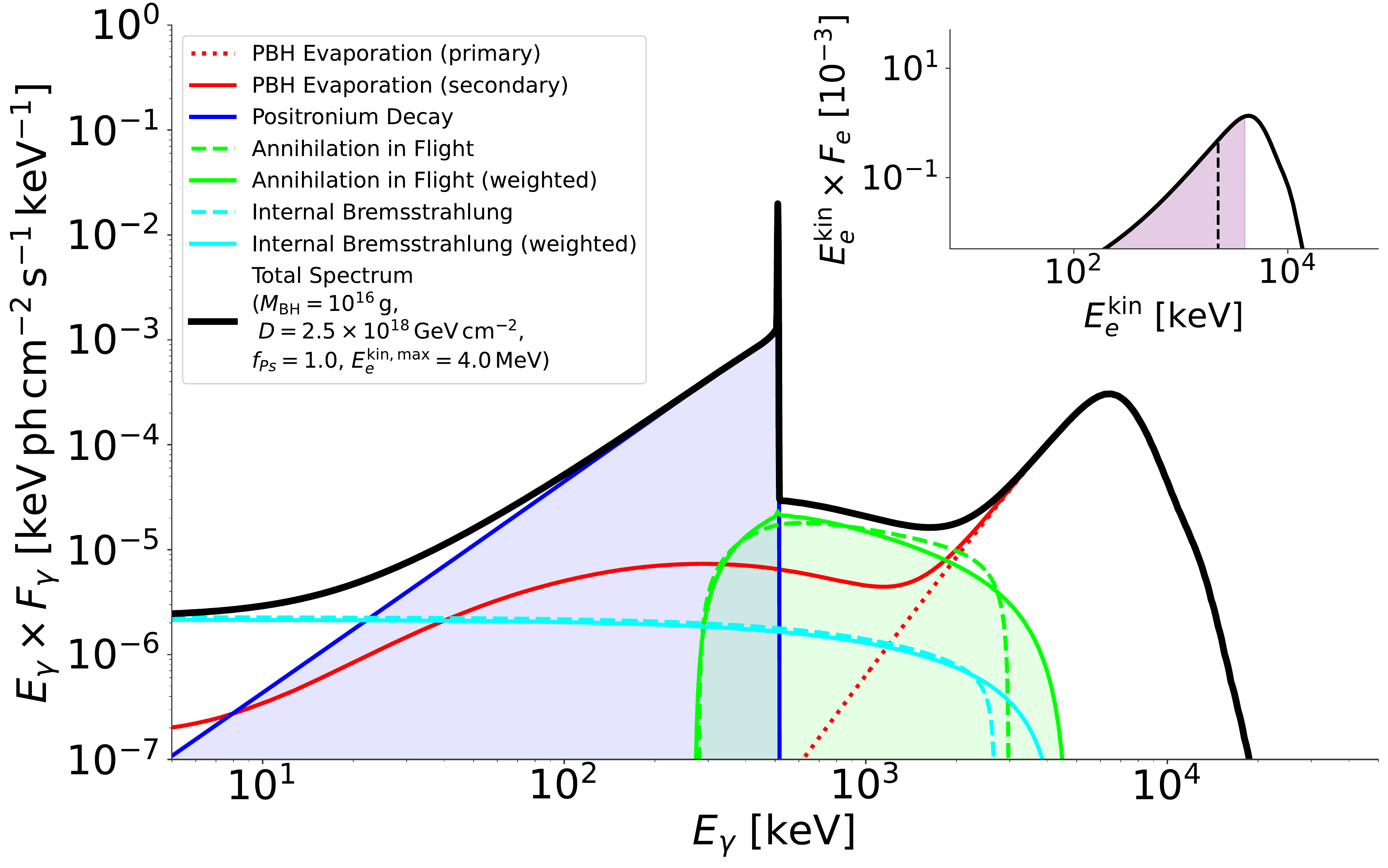}\\
	\includegraphics[width=\columnwidth,trim=0.0cm 0.0cm 0.0cm 0.0cm, clip=true]{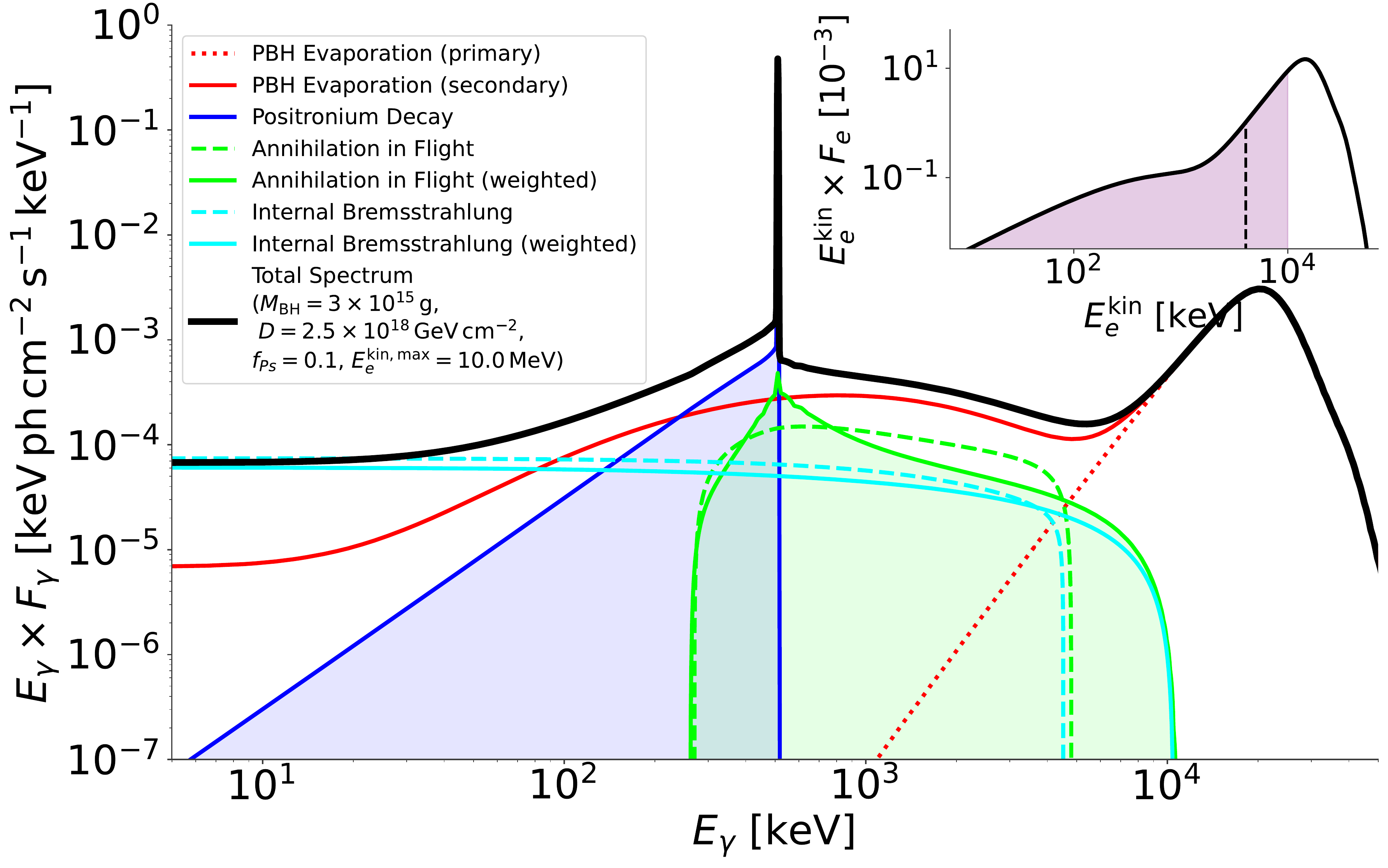}%
	\caption{Expected $\gamma$-ray spectra from PBH evaporation and subsequent annihilation of positron-electron pairs, plus internal bremsstrahlung of emitted pairs. The two panels show different model parameters (see legends) for a D-factor of $2.5 \times 10^{18}\,\mathrm{GeV\,cm^{-2}}$, consistent with Reticulum II. The small insets show the distribution of electrons and positrons from evaporation, together with the maximum kinetic energy $E_{\rm kin}^{\rm max}$ considered by the shaded area, and the mean kinetic energy of the electron/positron population (dashed line). The shaded area of the electron/positron distribution corresponds to the shaded areas in the $\gamma$-ray spectra.\label{fig:PBH_expectation}}
\end{figure}

The production of $e^+$s from BH evaporation has no kinematic threshold based on the mass of the BG.
However, the number of $e^+$s produced that may annihilate subsequently can lead to a 20--8000\,keV flux stronger than the evaporation signal itself which is given by a BH mass of $\lesssim 1.1 \times 10^{17}$\,g.
Given the particle spectrum of $e^+$s from \verb|BlackHawk|, the differential $e^+$ flux in a galaxy with D-factor $D$ is given by
\begin{equation}
	\frac{dF_e}{dE_e} = \frac{1}{4 \pi M_{\rm BH}} \frac{dN_e}{dE_e} D\mathrm{,}
	\label{eq:positron_flux}
\end{equation}
so that the total possible $e^+e^-$-annihilation flux $F_{\rm Ann}$ in that galaxy is
\begin{equation}
	F_{\rm Ann} = \int_{0}^{E_{\rm kin}^{\rm max}}\,dE_{\rm kin} \frac{dF_e}{dE_{e,\rm kin}}\,\mathrm{.}
	\label{eq:total_ann_flux}
\end{equation}
\noindent Here, $E_{\rm kin}^{\rm max}$ is the maximum kinetic energy of a $e^+$ that annihilates within the galaxy and does not escape into the intergalactic medium.
For the MW, $E_{\rm kin}^{\rm max}$ has been estimated to be around 3--7\,MeV \citep{Beacom2006_511,Sizun2006_511}.
We caution, however, that these estimates may be inaccurate because the statistical methods to compare the expected spectrum with the measurements of two different instruments are not rigorously correct.
Therefore, we leave $E_{\rm kin}^{\rm max}$ as a free parameter in our spectral fits.
Because the transport and environmental conditions in Ret\,II are largely unknown, this procedure effectively takes into account these uncertainties.
For reference, we discuss the two extreme cases of no and maximal $e^+$ annihilation in our results, Sec.\,\ref{sec:results}, together with the intermediate solution of marginalising over $E_{\rm kin}^{\rm max}$.

The total annihilation flux is composed of three components,
\begin{equation}
	F_{\rm Ann} = F_{511} + F_{\rm oPs} + F_{\rm IA}\mathrm{,}
	\label{eq:FAnn_three_comps}
\end{equation}
\noindent where $F_{511}$ is the flux in the 511\,keV line due to direct annihilation with $e^-$s and intermediate formation of Positronium (Ps), $F_{\rm oPs}$ is the three-photon decay flux of ortho-Ps, and $F_{\rm IA}$ is the total flux of $e^+$s annihilating in flight (IA) before stopping/thermalising.

The ortho-Ps and the line flux are related by quantum statistics as
\begin{equation}
	F_{\rm oPs} = \frac{9 f_{\rm Ps}}{8 - 6 f_{\rm Ps}} F_{511} =: r_{32} F_{511}\mathrm{,}
	\label{eq:FoPs_F511}
\end{equation}
\noindent with $f_{\rm Ps}$ being the Ps fraction, that is, the number of $e^+$s forming Ps before annihilating \citep{Leventhal1978_511}.
The scaling factor $r_{32}$ ranges between $0$ ($f_{\rm Ps} = 0$) and $4.5$ ($f_{\rm Ps} = 1$).
In general $f_{\rm Ps}$ depends on the annihilation conditions, for example the gas in which $e^+$s annihilate, which is a function of the temperature, density, and ionisation fraction, among others \citep[e.g.,][]{Jean2006_511,Churazov2005_511,Churazov2011_511,Siegert2016_511}.
Since the true conditions in Ret\,II are unknown, we leave $f_{\rm Ps}$ as free parameters in our spectral fits.
The ortho-Ps spectrum $\left(\frac{dN_{\gamma}}{dE_{\gamma}}\right)_{\rm oPs}$ has been calculated by \citet{Ore1949_511}, and we model the 511\,keV line, $\left(\frac{dN_{\gamma}}{dE_{\gamma}}\right)_{511}$ as a Gaussian at 511\,keV with a width according to SPI's energy resolution \citep{Diehl2018_BGRDB,Attie2003_SPI}.

\citet{Beacom2006_511} derived a relation between $F_{\rm IA}$ and the 511\,keV line flux,
\begin{equation}
	F_{\rm IA} = \frac{1}{1-\frac{3}{4}f_{\rm Ps}} \frac{1-P}{P} F_{511} =: r_{\rm IA} F_{511} \mathrm{,}
	\label{eq:FIA_F511}
\end{equation}
\noindent where $P = P(E_{\rm kin},m_e)$ is the probability of a $e^+$ with initial kinetic energy $E_{\rm kin}$ to annihilate before stopping/thermalising.
Therefore, the total annihilation flux can be expressed as
\begin{equation}
	F_{\rm Ann} = (1 + r_{32} + r_{\rm IA}) F_{511}\mathrm{.}
	\label{eq:FAnn_total}
\end{equation}

\noindent The survival probability in Eq.\,(\ref{eq:FIA_F511}) is given in general by
\begin{equation}
	P(E_{\rm kin}, m_e) = \exp\left(-n_{X} \int_{m_e}^{E_{\rm kin}}\,dE \frac{\sigma(E)}{\left|\frac{dE(E,n_X)}{dx}\right|}\right)\,\mathrm{,}
	\label{eq:P_surv}
\end{equation}
\noindent where $\sigma(E)$ is the total annihilation cross section, and $\left|\frac{dE(E,n_X)}{dx}\right|$ is the stopping power (energy loss rate, cooling function) of a $e^+$ propagating in a medium with density $n_X$.
For simplicity, we only consider Coulomb losses which are expected to dominate up to 100\,MeV for MW-like conditions, $\left|\frac{dE(E,n_X)}{dx}\right|_{\rm Coulomb} \propto n_X$, so that $n_X$ cancels.
The total IA spectrum for mono-energetic injection of $e^+$s thus reads
\begin{equation}
	\left(\frac{dF_{\gamma}}{dE_{\gamma}}\right)_{\rm IA}^{\rm mono} = F_{511} \frac{1}{1-\frac{3}{4}f_{\rm Ps}} \frac{1}{P} \left(\frac{dN_{\gamma}}{dE_{\gamma}}\right)_{\rm IA} \mathrm{,}
	\label{eq:IAspec_mono}
\end{equation}
\noindent with
\begin{equation}
	\left(\frac{dN_{\gamma}}{dE_{\gamma}}\right)_{\rm IA}^{\rm mono}  = \frac{n_X}{2 m_e} \int_{E_1}^{E_{\rm kin}}\,dE \frac{P(E_{\rm kin}, E) \frac{d\sigma\left(E,E_{\gamma}\right)}{dE}}{\left|\frac{dE(E,n_X)}{dx}\right|}  \mathrm{.}
	\label{eq:IAspectral_shape}
\end{equation}
where the integration limit $E_1 = \max\left( \frac{(E_{\gamma} -m_e)^2 + E_{\gamma}^2}{(E_{\gamma} - m_e) + E_{\gamma}} m_e, E_{\gamma} \right)$ is bound to obey energy conservation \citep{Svensson1982_ann_spec,Beacom2006_511}.
We note that $\int\,dE_{\gamma} \left(\frac{dN_{\gamma}}{dE_{\gamma}}\right)_{\rm IA} = 1 - P$.
Since in the case of PBHs, the injection energy is not mono-energetic, we calculate a weighted photon spectrum given the distribution of kinetic energies from \verb|BlackHawk| as
\begin{equation}
	\left(\frac{dN_{\gamma}}{dE_{\gamma}}\right)_{\rm IA} = \frac{ \int_{0}^{E_{\rm kin}^{\rm max}}\,dE_{\rm kin} \frac{dN_e}{dE_{e}} \left(\frac{dN_{\gamma}(E_{\rm kin})}{dE_{\gamma}}\right)_{\rm IA}^{\rm mono}  }{\int_{0}^{E_{\rm kin}^{\rm max}}\,dE_{\rm kin} \frac{dN_e}{dE_{e}}}\mathrm{.}
	\label{eq:fullIAspec}
\end{equation}

Radiative corrections in the production of $e^+e^-$-pairs can be described as internal bremsstrahlung (IB) which can be calculated without detailed knowledge of the previous process.
This final state radiation (FSR) arises as a result of the production itself and is independent of the astrophysical environment.
The photon source spectrum of IB per 511\,keV line photon is given by \citet{Beacom2005_DM_IB},
\begin{equation}
	\left(\frac{dF_{\gamma}}{dE_{\gamma}}\right)_{\rm IB}^{\rm mono} = \frac{1}{2}\frac{1}{1-\frac{3}{4}f_{\rm Ps}} \left(\frac{dN_{\gamma}}{dE_{\gamma}}\right)_{\rm IB}^{\rm mono}\mathrm{,}
	\label{eq:IB_source_spec}
\end{equation}
\noindent with
\begin{equation}
	\left(\frac{dN_{\gamma}}{dE_{\gamma}}\right)_{\rm IB}^{\rm mono} = \frac{\alpha}{2\pi} \left[\frac{M^2 + (M-2E_{\gamma})^2}{M^2 E_{\gamma}} \ln\left(\frac{M(M-2E_{\gamma})}{m_e^2}\right)\right]\mathrm{,}
	\label{eq:IB_photon_spec}
\end{equation}
\noindent where here, $M = E_{\rm kin} + m_e$ is the particles' injection energy from PBH evaporation.
Because the $e^+$-injection spectrum is again not mono-energetic, we weight over expected energy distribution such that
\begin{equation}
	\left(\frac{dN_{\gamma}}{dE_{\gamma}}\right)_{\rm IB} = \frac{ \int_{0}^{E_{\rm kin}^{\rm max}}\,dE_{\rm kin} \frac{dN_e}{dE_{e}} \left(\frac{dN_{\gamma}(E_{\rm kin})}{dE_{\gamma}}\right)_{\rm IB}^{\rm mono}} {\int_{0}^{E_{\rm kin}^{\rm max}}\,dE_{\rm kin} \frac{dN_e}{dE_{e}}}\mathrm{.}
	\label{eq:fullIAspec_weighted}
\end{equation}
\begin{figure}
	\centering
	\includegraphics[width=\columnwidth,trim=0.0cm 0.0cm 0.0cm 0.0cm, clip=true]{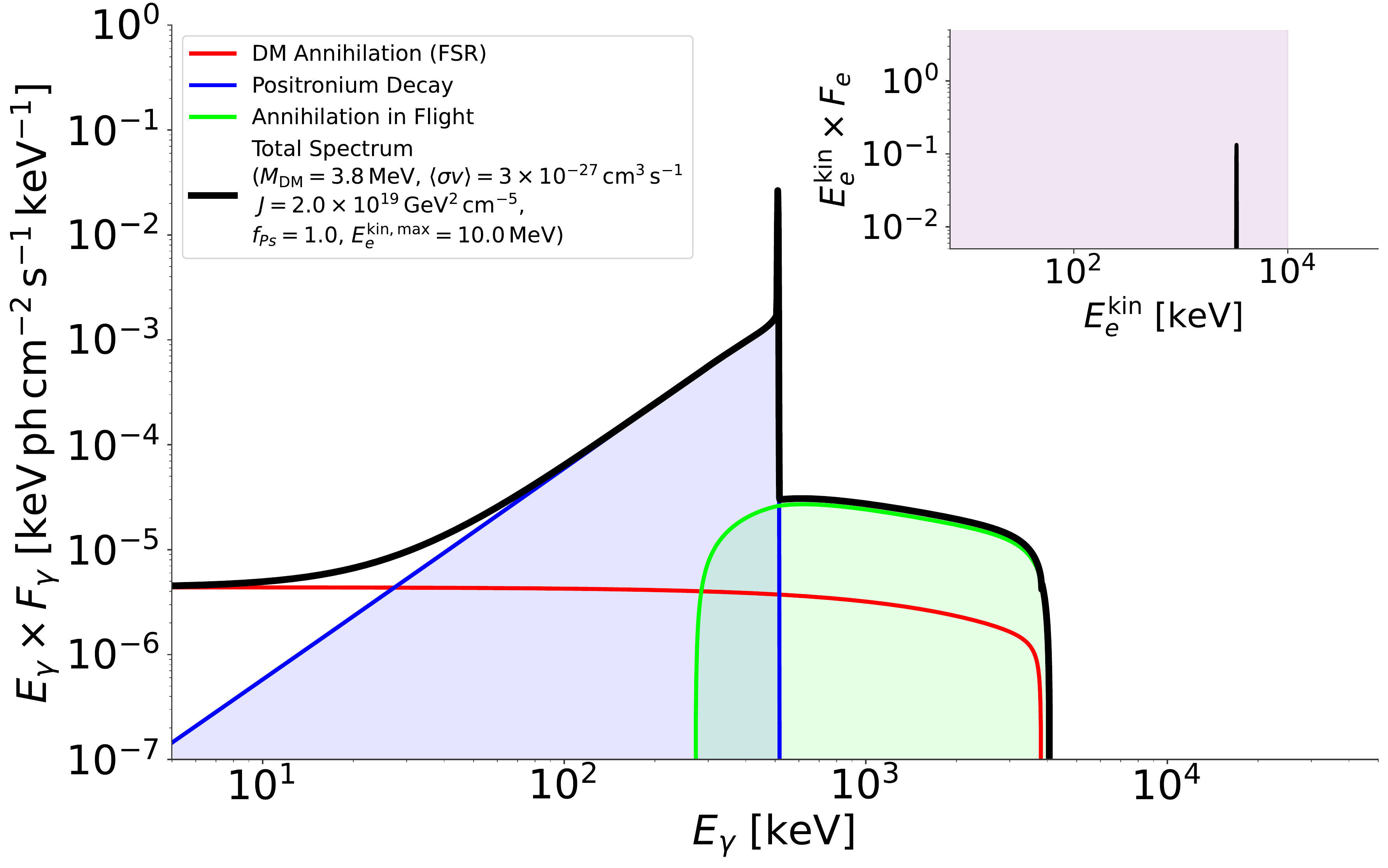}\\
	\includegraphics[width=\columnwidth,trim=0.0cm 0.0cm 0.0cm 0.0cm, clip=true]{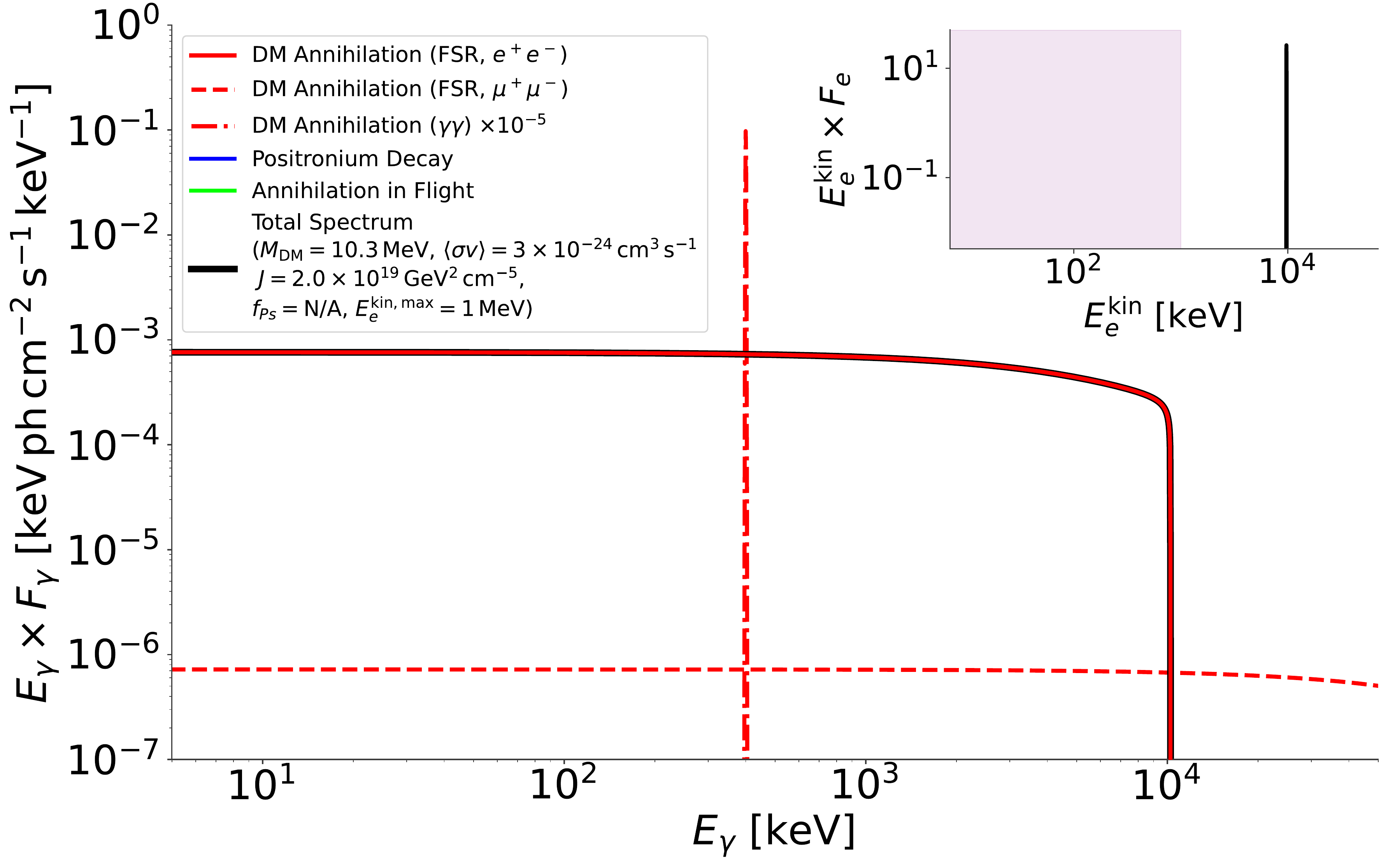}%
	\caption{Expected $\gamma$-ray spectra from light dark matter annihilation and subsequent annihilation of positron-electron pairs, similar to Fig.\,\ref{fig:PBH_expectation}. The two panels show different model parameters (see legends) for a J-factor of $2 \times 10^{19}\,\mathrm{GeV^2\,cm^{-5}}$. The lower panel also shows model expectations from final state radiation of the $\mu^+\mu^-$-channel and direct decay into two photons.\label{fig:DM_expectation}}
\end{figure}

\noindent Finally, the total photon source spectrum of PBH evaporation including subsequent $e^+$-annihilation and IB in a galaxy is the sum the components:
\begin{eqnarray}
	& \left(\frac{dF_{\gamma}}{dE_{\gamma}}\right)_{\rm PBH}^{\rm tot} = \left(\frac{dF_{\gamma}}{dE_{\gamma}}\right)_{\rm PBH} + \left(\frac{dF_{\gamma}}{dE_{\gamma}}\right)_{\rm IB} + \left(\frac{dF_{\gamma}}{dE_{\gamma}}\right)_{\rm Ann} = \nonumber\\
	& = \left(\frac{dF_{\gamma}}{dE_{\gamma}}\right)_{\rm PBH} + \left(\frac{dF_{\gamma}}{dE_{\gamma}}\right)_{\rm IB} + \sum_{i \in \mathscr{A}} \left(\frac{dF_{\gamma}}{dE_{\gamma}}\right)_{i} \mathrm{,}
	\label{eq:total_PBH_spec}
\end{eqnarray}
\noindent with $\mathscr{A} = \left\{511, \rm oPs, \rm IA\right\}$.
$\left(\frac{dF_{\gamma}}{dE_{\gamma}}\right)_{\rm PBH}$ is a function of $M_{\rm BH}$, $D$ and $f_{\rm PBH}$, and $\left(\frac{dF_{\gamma}}{dE_{\gamma}}\right)_{\rm Ann}$ is a function of $f_{\rm Ps}$ and $E_{\rm kin}^{\rm max}$.
We show two examples of the total emission spectrum from PBH evaporation in Ret\,II in Fig.\,\ref{fig:PBH_expectation}.

\begin{figure*}
	\centering
	\includegraphics[width=0.33\textwidth,trim=3.5cm 9.7cm 3cm 7.5cm, clip=true]{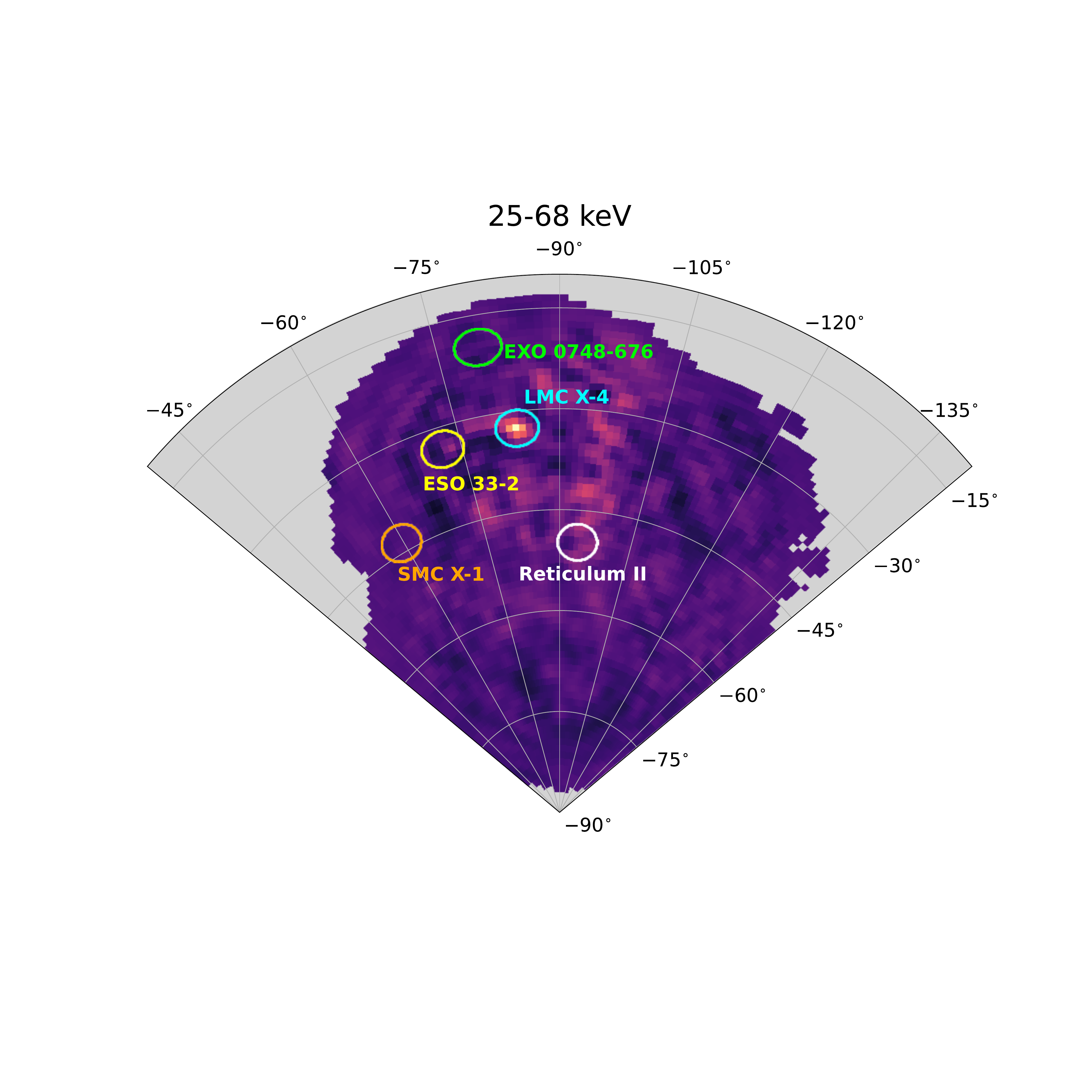}%
	\includegraphics[width=0.33\textwidth,trim=3.5cm 9.7cm 3cm 7.5cm, clip=true]{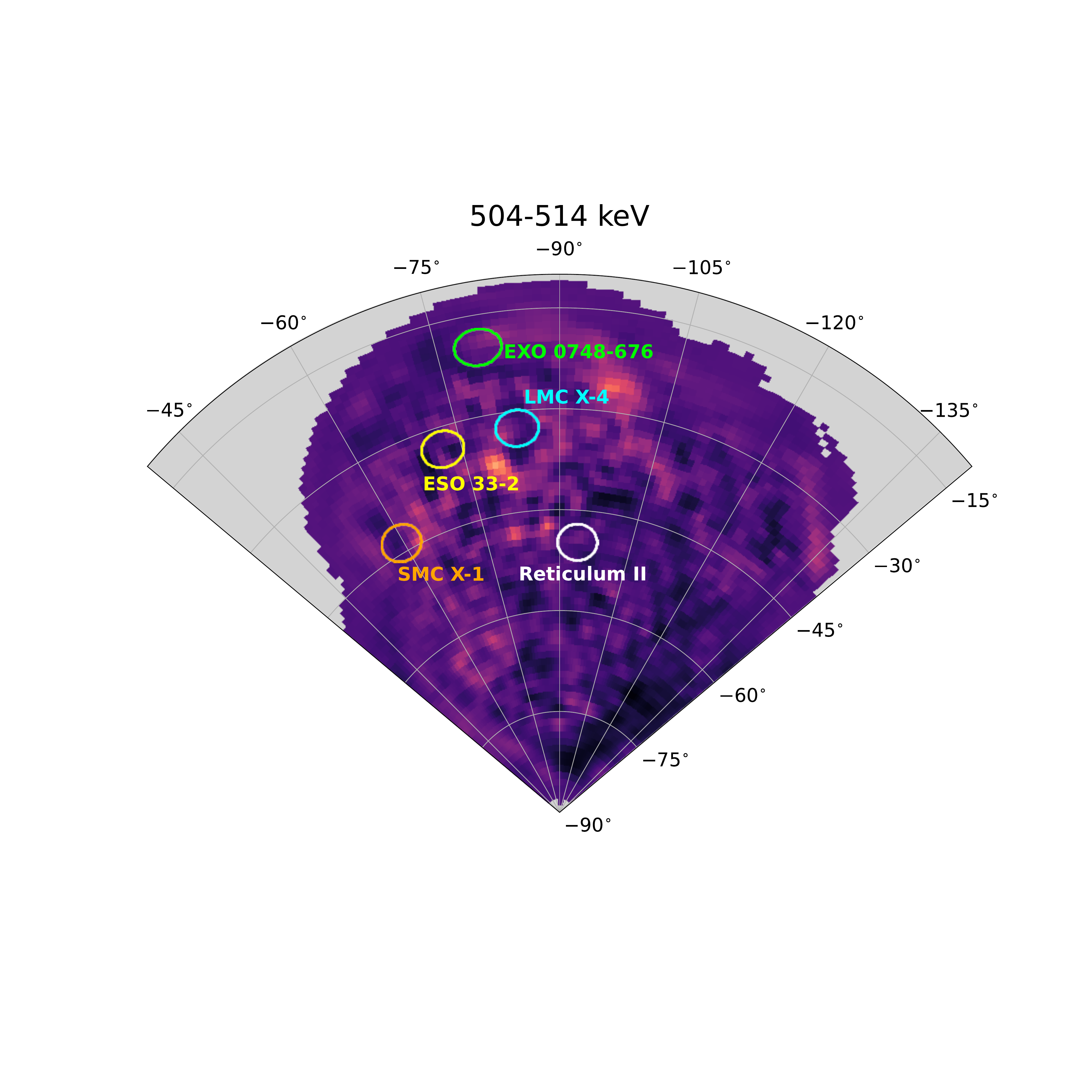}%
	\includegraphics[width=0.33\textwidth,trim=3.5cm 9.7cm 3cm 7.5cm, clip=true]{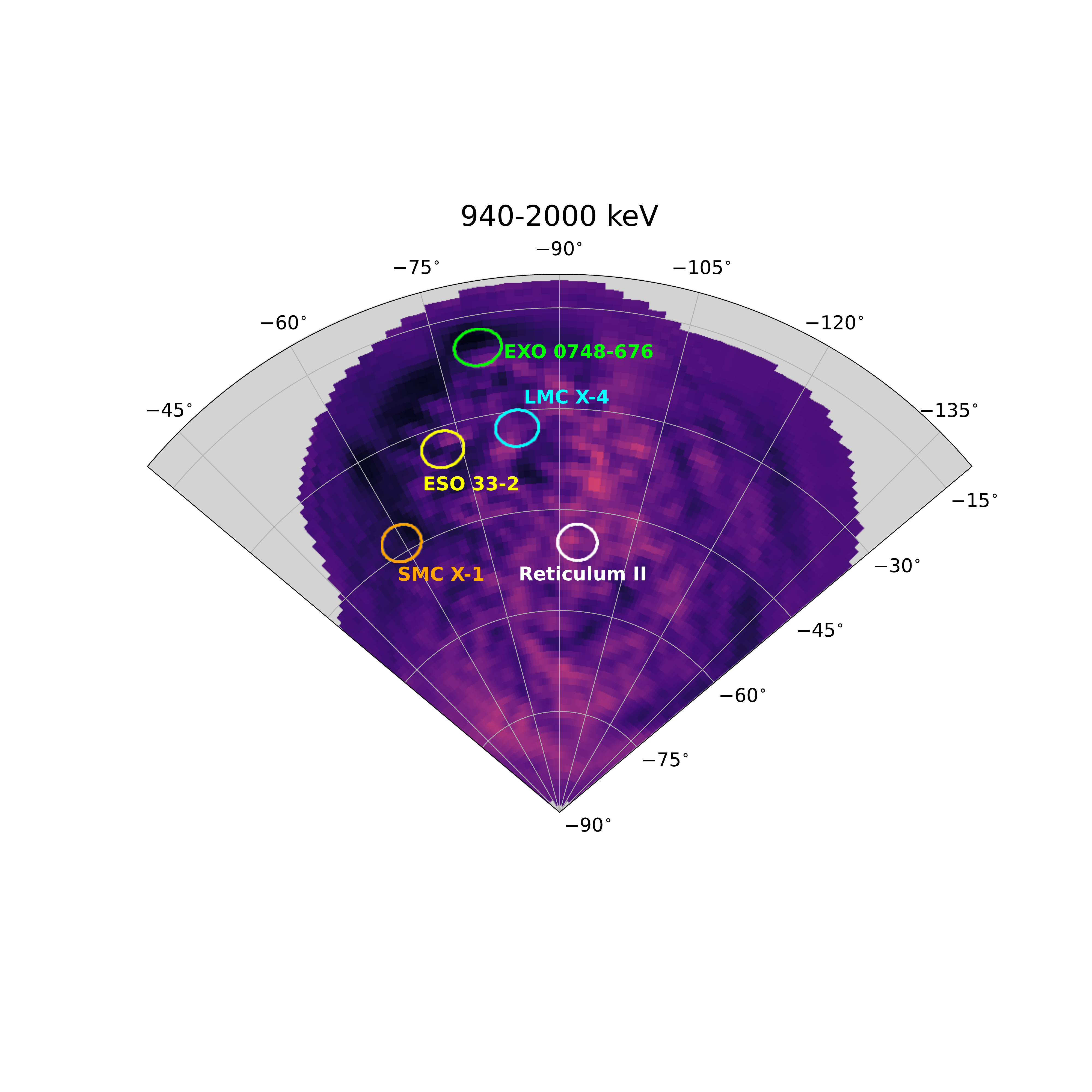}%
	\caption{Count residuals projected back to the the sky by applying the imaging response backwards. The radial and azimuthal coordinates are the Galactic latitude and longitude, respectively. Shown are the bands 25--68\,keV, 508--514\,keV, and 940--2000\,keV as examples. Only LMC X-4 is significantly detected in the first energy band.\label{fig:sky_residuals}}
\end{figure*}
\subsubsection{Particle Dark Matter Annihilation and Decay}\label{sec:WIMP_spectra}
We consider light ($m_{\rm DM} \lesssim 300\,\mathrm{MeV}$) DM particles that either annihilate or decay directly into standard model particles, leading to an expected photon emission spectrum.
Therefore we evaluate the cases
\begin{enumerate}
	\itemsep0em 
	\item $\mathrm{DM + DM \rightarrow e^+ + e^- + (\gamma)_{\rm FSR}}$,
	\item $\mathrm{DM + DM \rightarrow \gamma + \gamma}$,
	\item $\mathrm{DM \rightarrow e^+ + e^- + (\gamma)_{\rm FSR}}$,
	\item \begin{enumerate}
		\item $\mathrm{DM \rightarrow \gamma + \gamma}$, and
		\item $\mathrm{DM \rightarrow Y + \gamma}$,
	\end{enumerate}
\end{enumerate}
\noindent where $\mrm{Y}$ is any relativistic particle, for example a sterile neutrino that decays into $\gamma + \nu$.
The photon spectra of the FSR in cases 1. and 3. are identical to the IB spectrum,
\begin{equation}
	\left(\frac{dN_{\gamma}}{dE_{\gamma}}\right)_{\rm FSR} \equiv \left(\frac{dN_{\gamma}}{dE_{\gamma}}\right)_{\rm IB}\mathrm{,}
	\label{eq:IBequivFSR}
\end{equation}
\noindent however now, $M = 2m_{\rm DM}$ for annihilation and $M = 1m_{\rm DM}$ for decay.
Another factor of $\frac{1}{2}$ is required in Eq.\,(\ref{eq:IB_photon_spec}) for Dirac DM annihilation/decay to account for distinct particle-antiparticle DM with equal densities.
The expected photon spectrum of direct annihilation/decay into two photons (cases 2. and 4.) is a delta-function,
\begin{equation}
	\left(\frac{dN_{\gamma}}{dE_{\gamma}}\right)_{\gamma\gamma} = 2\delta\left(E_{\gamma} - \frac{M}{2}\right)\mathrm{.}
	\label{eq:delta_function_spectrum}
\end{equation}
Case 4. (b) differs from case 4. (a) only by an additional factor of $2$.
We model Eq.\,(\ref{eq:delta_function_spectrum}) as Gaussians with the energy resolution of SPI.

Similar to the PBH case, the emitted $e^+$s from DM particles in a galaxy's halo might annihilate during (IA) and after (Ps) propagation in the galaxy, or escape into the intergalactic medium if their injection energy is larger than a threshold $E_{\rm kin}^{\rm max}$.
The only difference is now that the energy distribution of produced pairs is mono-energetic, so that either all or none of the pairs lead to additional $e^+$-annihilation related spectra.
Given the equivalence in Eq.\,(\ref{eq:IBequivFSR}), one can estimate the total 511\,keV line flux from DM particle annihilation as
\begin{equation}
	F_{511}^{\rm Ann} = \left(1 - \frac{3}{4}f_{\rm Ps}\right) \frac{J \langle \sigma v \rangle}{2 \pi m_{\rm DM}^2}\mathrm{,}
	\label{eq:F511_from_DMAnn}
\end{equation}
\noindent and from DM particle decay as
\begin{equation}
	F_{511}^{\rm Dec} = \left(1 - \frac{3}{4}f_{\rm Ps}\right) \frac{D}{2 \pi m_{\rm DM} \tau}\mathrm{.}
	\label{eq:F511_from_DMDecay}
\end{equation}
The remaining spectral components from ortho-Ps and annihilation in flight are the same as in Sec.\,\ref{sec:PBH_evaporation}, except that now the mono-energetic cases are considered.
The total source photon spectrum is again the sum of the components, here FSR, 511\,keV line, ortho-Ps, and in-flight annihilation.
We show examples of the expected DM annihilation spectra in Fig.\,\ref{fig:DM_expectation}.

\section{Results}\label{sec:results}
We examine our model fits to measured data through a careful analysis of the residuals in different data space dimensions.
Herein, it proves valuable to apply a back-projection onto the sky of the count residuals.
For cases of background model imperfections, we would discover irregular or large-scale patterns on the sky, whereas for imperfections of the sky model (e.g., missing a point source) would show up as a single peak at the source's location.
In Fig.\,\ref{fig:sky_residuals}, we show the residual count map of an instrumental-background-only fit in three example energy bands.
Only the high-mass X-ray binary LMC\,X-4 is detected with a significance of $21\sigma$ up to an energy of $\sim 200$\,keV.
The other sources, and in particular Ret\,II, are not seen in any energy band.
The count residuals show no strong patterns as expected from an otherwise empty field.
In the 511\,keV band, the residuals appear more structured, which is a result of the reduced count rate compared to other bands.
For reference, the particularly bright spot between ESO\,33-2 and LMC\,X-4 at $(l,b) = (-78^{\circ},-37^{\circ})$ has a significance of $\sim 2.5\sigma$, and is thus considered a background fluctuation.
Details about the 511\,keV line from Ret\,II are given in Sec.\,\ref{sec:511keV_line}.
\begin{figure}
	\centering
	\includegraphics[width=\columnwidth,trim=0.0cm 0.2cm 0.9cm 0.0cm,clip=true]{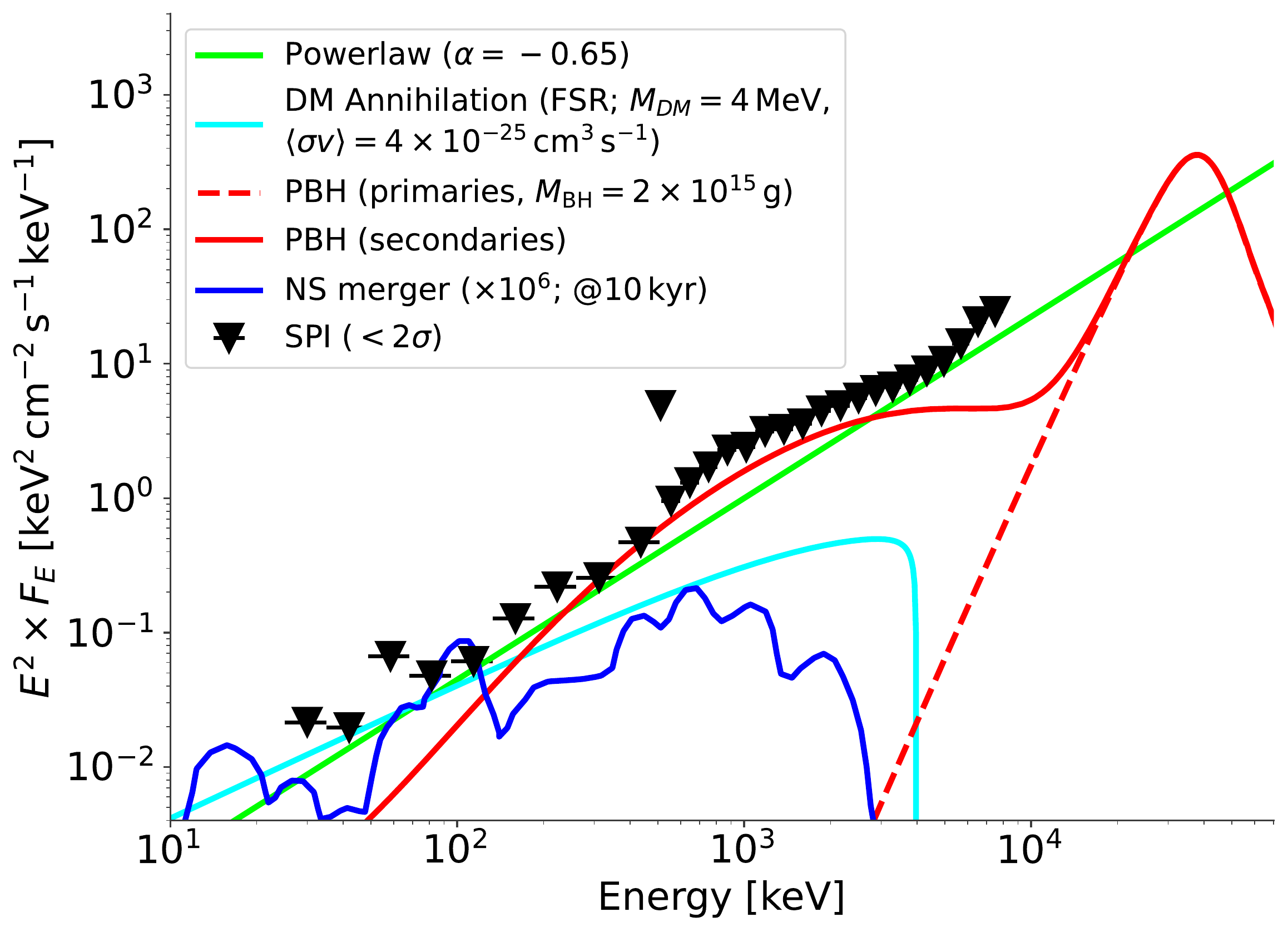}%
	\caption{SPI spectrum from the position of Reticulum II. Shown are $2\sigma$ upper limits on the flux as a function of photon energy. A selection of excluded models is shown (see below).\label{fig:spectrum}}
\end{figure}

\subsection{Total Spectrum}\label{sec:total_spec}
We show the spectrum measured by SPI from the direction of Ret\,II, spatially modelled as a point source, Eq.\,(\ref{eq:point_source}), in Fig.\,\ref{fig:spectrum}.
Because the source is not detected in any of the energy bins, we plot $2\sigma$ upper limits.
For comparison, we show several models that are excluded, given these fluxes.
Detailed exclusion plots are found in the following sections.
From a spectral fit with a generic power-law, $\propto \left(\frac{E_{\gamma}}{1\mathrm{MeV}}\right)^{\nu}$, we derive an upper bound on the total flux in the band 20--8000\,keV of $10^{-8}\,\mathrm{erg\,cm^{-2}\,s^{-1}}$.

\subsection{511 keV Line}\label{sec:511keV_line}
We find a 511\,keV line flux limit of $1.7 \times 10^{-4}\,\mathrm{ph\,cm^{-2}\,s^{-1}}$.
This value is about one order of magnitude smaller than the $3\sigma$ signal reported by \citet{Siegert2016_dsph} of $(1.7 \pm 0.5) \times 10^{-3}\,\mathrm{ph\,cm^{-2}\,s^{-1}}$.
Given the enhanced exposure time and the source being inside the FCFOV all the time thanks to the new dedicated observation campaign, the improvement in sensitivity is plausible.

\begin{figure}
	\centering
	\includegraphics[width=\columnwidth,trim=1.0cm 0.0cm 3.5cm 2.2cm,clip=true]{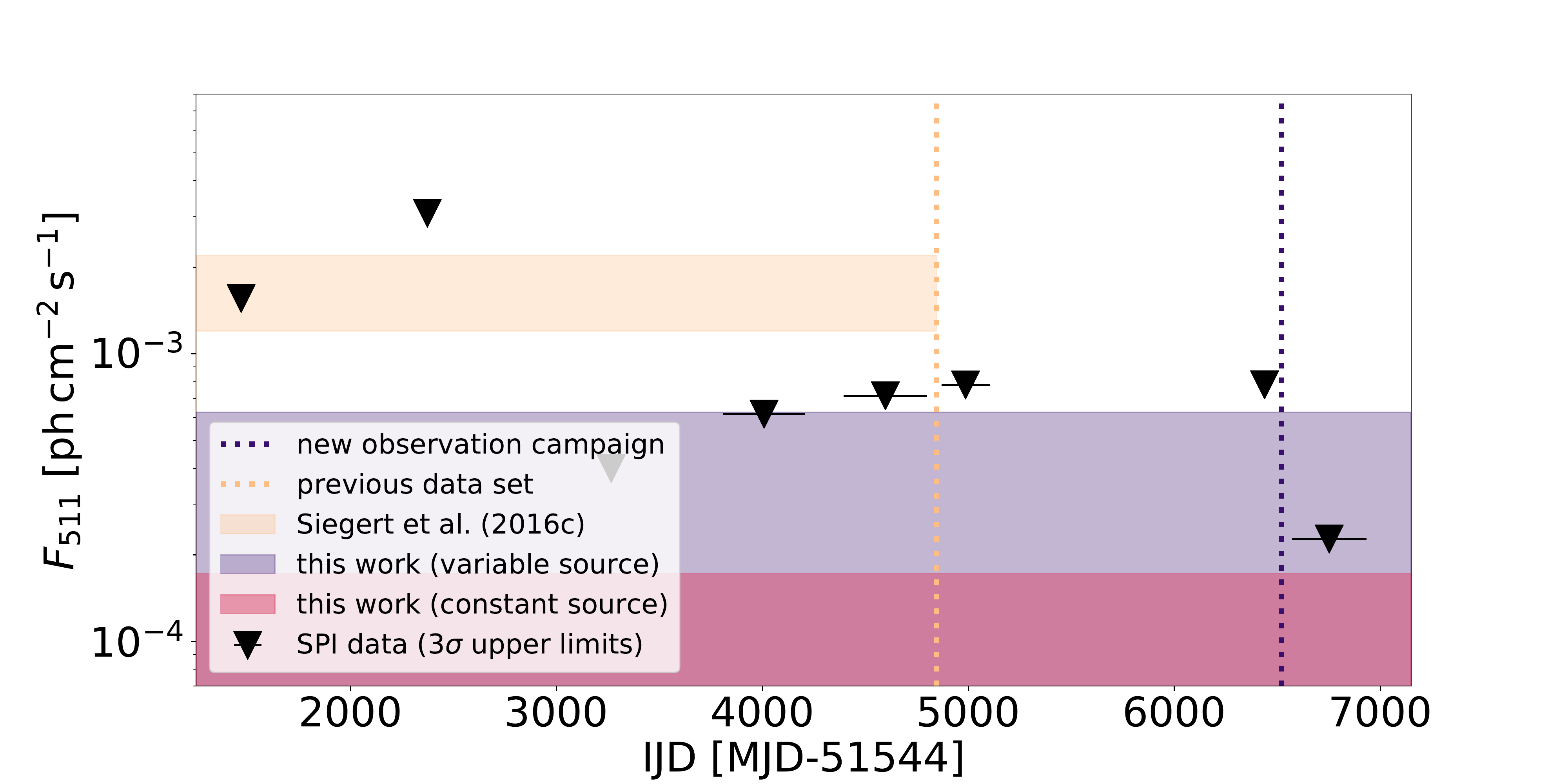}%
	\caption{Lightcurve of the 511\,keV line from Ret\,II. No positive excess is found at any time. Therefore, shown are $3\sigma$ upper limits in the band 508--514\,keV. The inferred flux from \citet{Siegert2016_dsph} is shown in comparison to the new INTEGRAL observation campaign. Additional serendipitous observations between IJD 4800 and 6600 are also included.\label{fig:lc511}}
\end{figure}

The previous detection of Ret\,II might be due to an intrinsic variability in time, as could be expected from the outburst of a microquasar \citep{Siegert2016_V404} or a NS merger \citep{Fuller2018_511NS}, for example.
We therefore split the data set into different time bins according to chance observations of nearby targets as well as the new observation campaign.
The resulting 511\,keV light curve is shown in Fig.\,\ref{fig:lc511}.
Considering only observations in which Ret\,II is inside the FCFOV results in upper limits consistent with the previous measurement by \citet{Siegert2016_dsph}.
We find that the significance of the 511\,keV line rises up to $\sim 2\sigma$ until the end of the older data set.
Including more data after 2013 and in particular the additional 1\,Ms in 2018 ($\mathrm{IJD} \sim 6700$), leads to a greatly reduced upper flux bound as well as a lower significance.
Assuming the source to be variable in time, the upper limit on the 511\,keV line is $\sim 6.1 \times 10^{-4}\,\mathrm{ph\,cm^{-2}\,s^{-1}}$ -- barely consistent with the previous estimate.
We conclude that the $3\sigma$ signal from Ret\,II by \citet{Siegert2016_dsph} was most likely due to an instrumental background fluctuation, paired with short exposures outside the FCFOV.
We can not, however, exclude that the previously reported signal was of astrophysical origin.

\subsection{Primordial Black Holes}\label{sec:PBHs}
\begin{figure}
	\centering
	\includegraphics[width=\columnwidth,trim=0.0cm 0.0cm 0.0cm 0.0cm, clip=true]{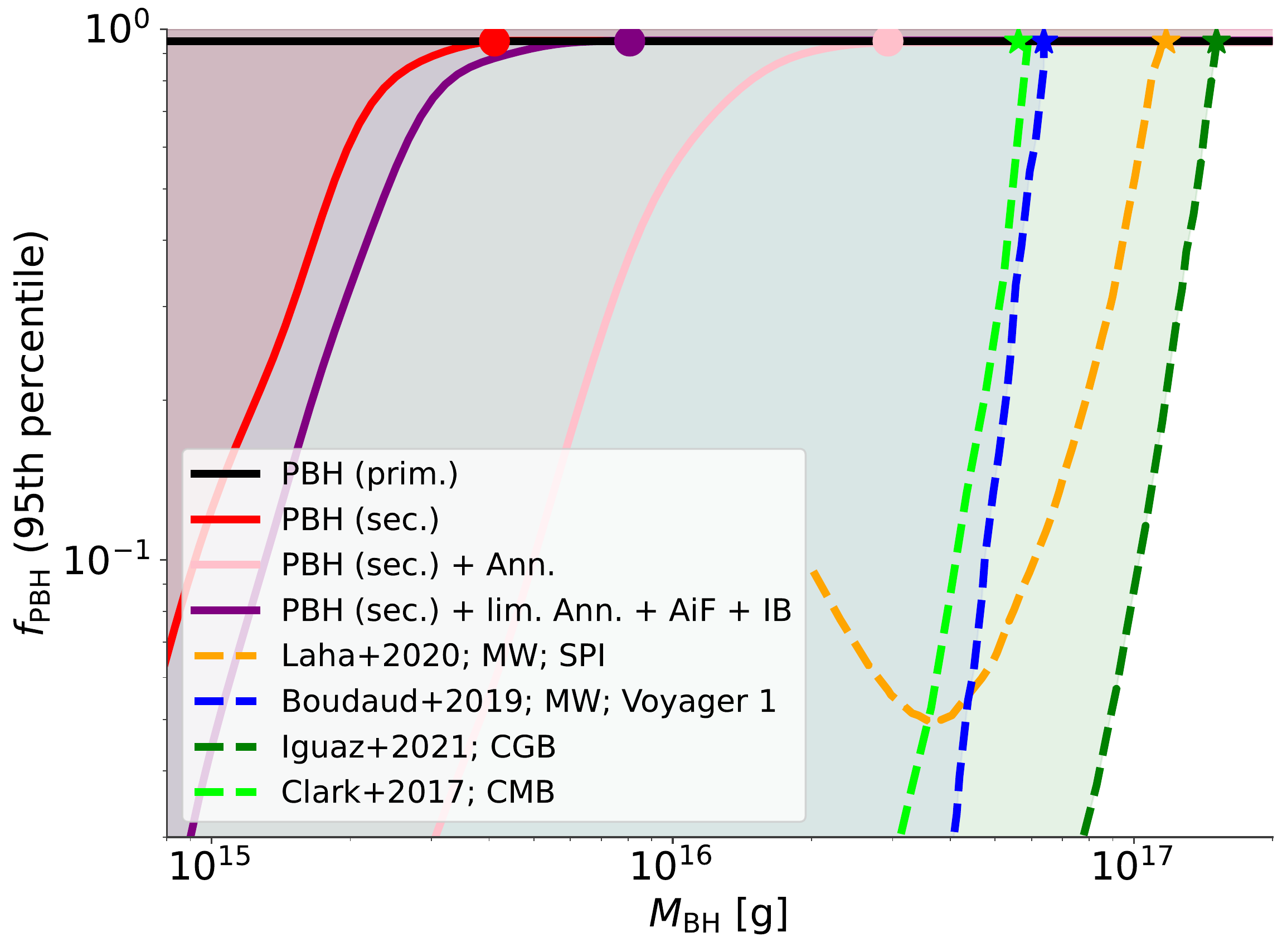}%
	\caption{Fraction of DM that can be constituted of a monochromatic PBH distribution with mass $M_{\rm BH}$ in Ret\,II (solid lines), compared to literature estimates. The circles mark the masses when the 95th percentile of the posterior $\pi(f_{\rm PBH},M_{\rm BH}|y_i,\sigma_i)$ in each case is reached. Stars mark the literature limits of 100\,\% PBH DM. \label{fig:PBH_limits}}
\end{figure}

We perform fits to the extracted spectrum, Fig.\,\ref{fig:spectrum}, with Eq.\,(\ref{eq:total_PBH_spec}), and sub-cases thereof.
Here we emphasise that we do \emph{not} solve for the interesting parameters until a certain figure of merit is above a threshold, but determine the joint posterior distributions of all the free parameters in Eq.\,(\ref{eq:total_PBH_spec}).
That means we include the uncertainties on $D$ and estimate $f_{\rm PBH}$ conditioned on all the other unknown parameters, and illustrate the limits on $f_{\rm PBH}$ as a function of $M_{\rm BH}$, as it is typically done.
However here we consider the 95th percentile in $f_{\rm PBH}$-direction of the marginalised posterior $\pi(f_{\rm PBH},M_{\rm BH}|y_i,\sigma_i)$ with $y_i$ and $\sigma_i$ as the flux values and corresponding uncertainties of energy bin $i$ in the spectrum.
See Appendix\,\ref{sec:appendix_spec_fits} for details.
In Fig.\,\ref{fig:PBH_limits}, we show our upper bounds on $f_{\rm PBH}$, and compare them to estimates from the recent literature, including the MW \citep{Laha2019_511PMBHs,Laha2020_PMBHDM}, cosmic rays \citep{Boudaud2019_CR_Voyager_PBH}, the cosmic microwave background \citep[CMB,][]{Clark2017_CMBPBH}, and the cosmic $\gamma$-ray background \citep[CGB][]{Iguaz2021_CGBPBH}.

The primary component of PBH evaporation alone cannot constrain $f_{\rm PBH}$ in Ret\,II.
Including the secondary particles, the upper bound on $f_{\rm PBH} = 1$ excludes monochromatic PBH distributions with masses of $\lesssim 0.4 \times 10^{16}$\,g.
Assuming that all $e^+$s created by PBHs in Ret\,II annihilate strengthens the bound on $f_{\rm PBH}$, then excluding masses $\lesssim 2.9 \times 10^{16}$\,g.
Since this assumption is too optimistic, we included the limiting factor for $e^+$ annihilation, that all $e^+$s above the threshold $E_{\rm kin}^{\rm max}$ escape from the galaxy and do not contribute to the annihilation spectrum.
This describes our most realistic bound on $f_{\rm PBH} = 1$, which excludes masses $\lesssim 0.8 \times 10^{16}$\,g.

\citet{Laha2019_511PMBHs} performed a similar estimate for the flux of the 511\,keV line in the MW, however assumed that the entire annihilation flux, from integrating over the $e^+$-distribution of PBHs, ends up in the 511\,keV line.
Since the Ps fraction in the MW is within a range $0.97$--$1.00$, the 511\,keV line flux from PBHs is at least four times smaller, because most of the annihilation flux goes into the three-photon decay of ortho-Ps.
In return this means that the bounds on $f_{\rm PBH}$ from \citet{Laha2019_511PMBHs} should be shifted by about a factor of four `vertically', which decreases the mass at $f_{\rm PBH} = 1$ from $\sim 8.9 \times 10^{16}$\,g to $\sim 6.4 \times 10^{16}$\,g.
We test this by making the same assumptions with our Ret\,II spectrum, and find $f_{\rm PBH} = 1$ is excluded for masses $M_{\rm BH} \lesssim 4.0 \times 10^{16}$\,g.

Our conservative and realistic bounds are about one order of magnitude in PBH mass below previous estimates.
Nevertheless, our data set is considerably shorter than all the other data sets, often comprising more than a decade of observation time.
Thanks to the much smaller region of interest in the case of Ret\,II, basically constituting a point source of SPI, our systematic uncertainties can be considered smaller compared to studies of the MW.
The entire Galactic (DM) profile is rather uncertain, and also the Galactic centre itself bears problems when DM models are applied, as is often the case for diffuse emission in the GeV band.
While our bounds are not yet as constraining, we show that reasonable progress can be made with even a small data set.
In particular, we show that when all uncertainties on the spectral parameters are included also more conservative and realistic bounds can result.

\subsection{Particle Dark Matter}\label{sec:particle_DM}
Similar to the PBH case, we perform a fit to the extracted Ret\,II spectrum and marginalise over the uncertain parameters to determine the posteriors $\pi(\langle \sigma v \rangle, m_{\rm DM}| y_i, \sigma_i)$ and $\pi(\tau, m_{\rm DM} | y_i, \sigma_i)$ for DM annihilation and decay, respectively.
In what follows, the excluded regions consider the 95th percentile of the posteriors in $\langle \sigma v \rangle$- and $\tau$-direction, respectively.
\begin{figure}
	\centering
	\includegraphics[width=\columnwidth,trim=0.0cm 0.0cm 0.0cm 0.0cm, clip=true]{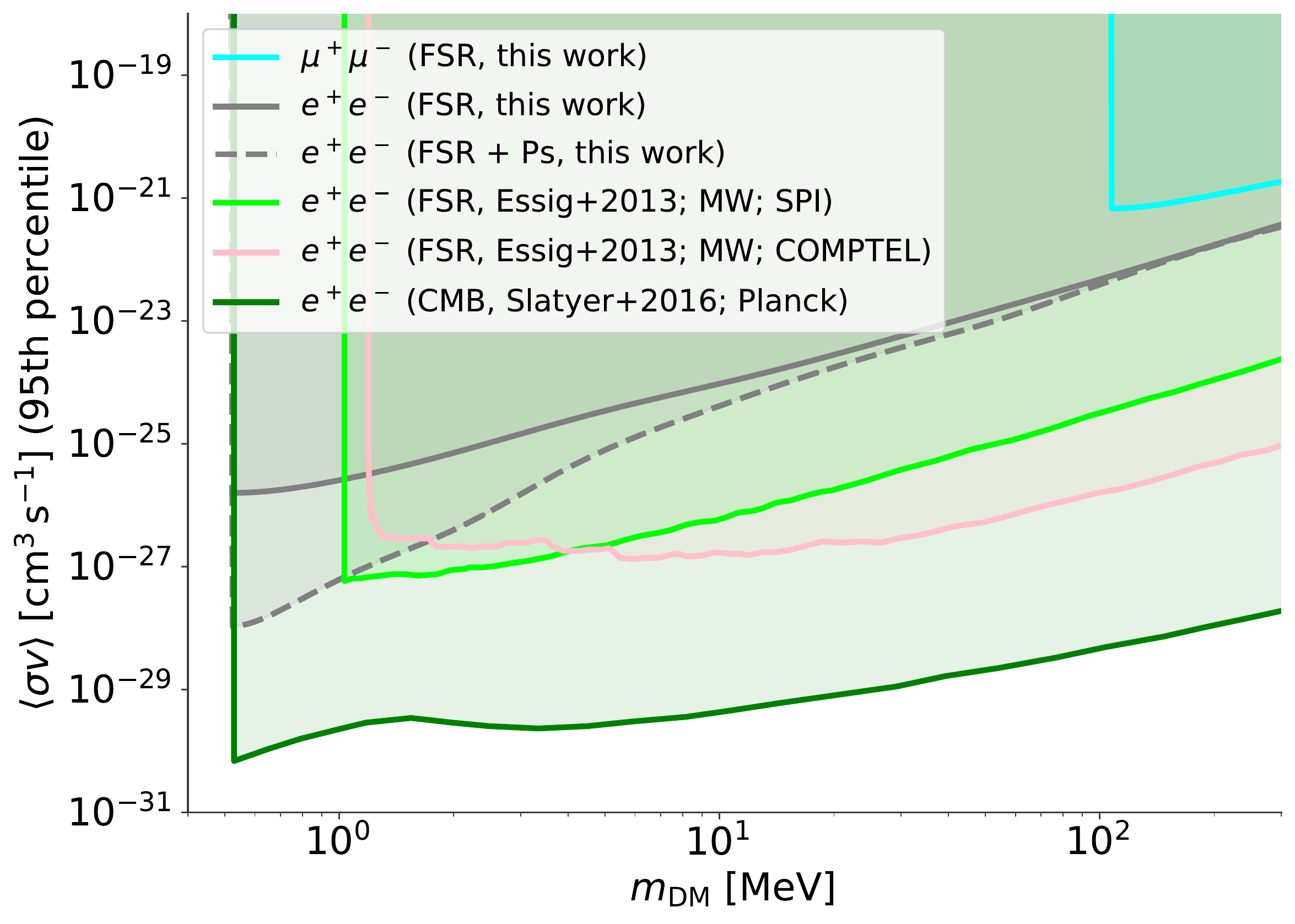}%
	\caption{DM annihilation cross section into $e^+e^-$ with subsequent FSR and possible annihilation of the pairs in Ret\,II (case 1. in Sec.\,\ref{sec:WIMP_spectra}; gray lines), compared to literature limits. The shaded regions are excluded. \label{fig:DMAnn_FSR_limits}}
\end{figure}
\begin{figure}
	\centering
	\includegraphics[width=\columnwidth,trim=0.0cm 0.0cm 0.0cm 0.0cm, clip=true]{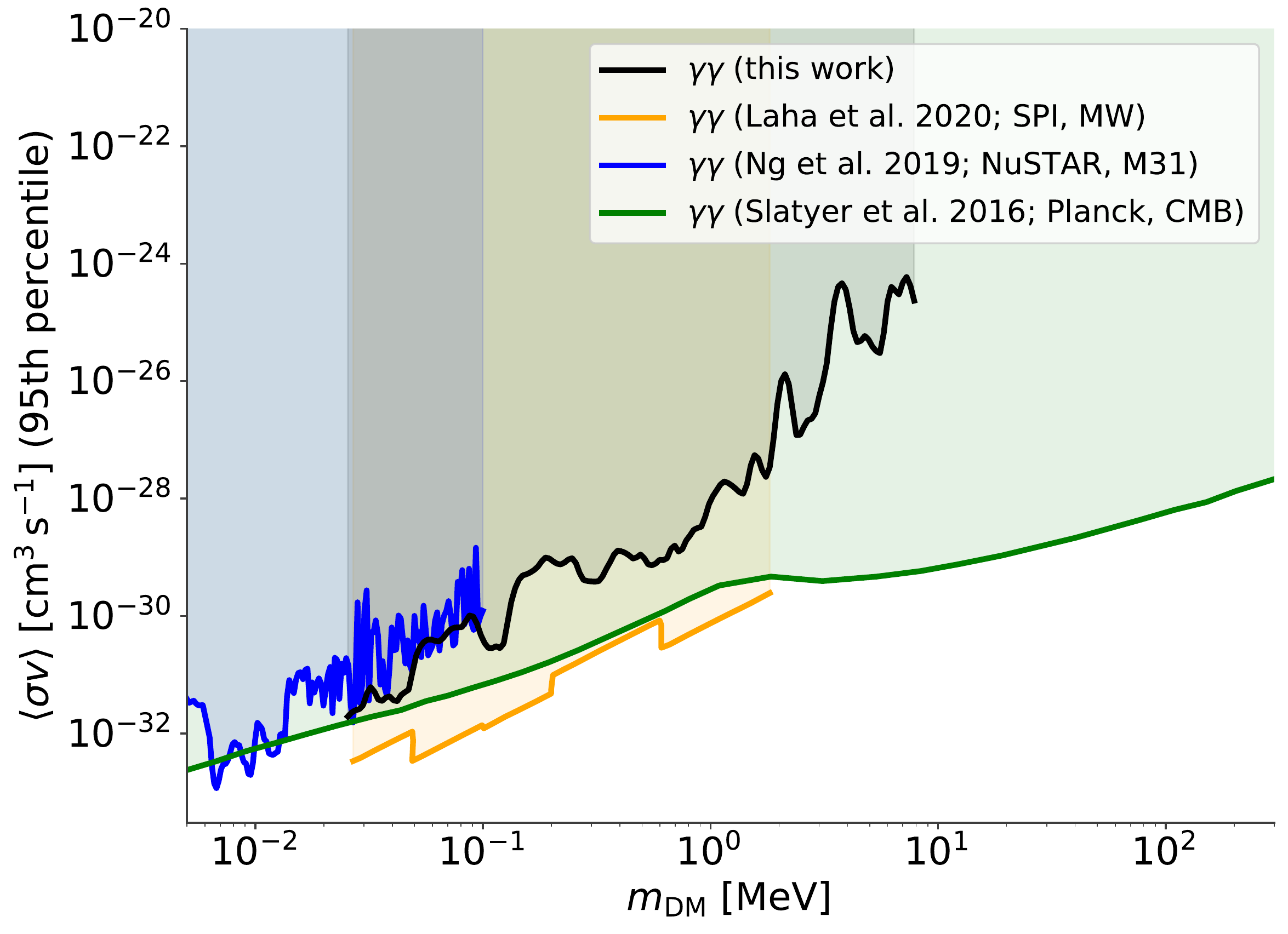}%
	\caption{Similar to Fig.\,\ref{fig:DMAnn_FSR_limits} but for $\mathrm{DM + DM \rightarrow \gamma + \gamma}$.\label{fig:DMAnn_gammagamma_limits}}
\end{figure}
\subsubsection{Annihilation}
In Fig.\,\ref{fig:DMAnn_FSR_limits}, we show the excluded region of the velocity-averaged DM annihilation cross section into $e^+e^-$, followed by FSR and subsequent $e^+$ annihilation.
Without additional $e^+$ annihilation, the bounds from Ret\,II are barely comparable to literature values from other $\gamma$-ray experiments \citep{Essig2013_DMlimits_gammarays} or the CMB \citep{Slatyer2016_CMB_DM}.
When limited $e^+$ annihilation is included, we can improve the limits in the DM mass range $0.5$--$1.0\,\mathrm{MeV}$, and find a general trend of the annihilation cross section at tree level of $\langle \sigma v \rangle \lesssim 5 \times 10^{-28} \left(m_{\rm DM} / \mathrm{MeV} \right)^{2.5}\,\mathrm{cm^3\,s^{-1}}$, based on the shape of the measured limits as a function of energy in Fig.\,\ref{fig:DMAnn_FSR_limits}.
For completeness, we show the limits on DM annihilation into $\mu^+\mu^-$, being hardly constrained by Ret\,II data.

The cross section limit for the case $\mathrm{DM + DM \rightarrow \gamma + \gamma}$ is shown in Fig.\,\ref{fig:DMAnn_gammagamma_limits}.
Our limits on DM annihilation into two photons are comparable to those from NuSTAR observations of M31 \citep{Ng2019_DM_NuSTAR_M31}, and about one order of magnitude worse than from MW observations with SPI \citep{Laha2020_PMBHDM}.
Because our data set extends to the full energy range of SPI, we can set however limits out to $m_{\rm DM} \leq 8\,\mathrm{MeV}$, varying between $10^{-28}$ and $10^{-25}\,\mathrm{cm^3\,s^{-1}}$.
However, limits from the CMB \citep{Slatyer2016_CMB_DM} are still more stringent at there energies.

\subsubsection{Decay}
The properties of decaying DM particles cannot be inferred directly from the previous case of annihilating DM.
First, the kinematic thresholds for standard model particle production is shifted by a factor of two which leads to altered (shifted) spectral shapes in particular for FSR and IA.
Second, the J-factor of Ret\,II is about twice as uncertain as its D-factor, which allows to populate different regions in the full posterior.
We therefore perform spectral fits for the DM decay case separately to infer bounds on the DM lifetime $\tau$.
This also results in slightly worse bounds when considering decay compared to annihilation (cf. NuSTAR limits).
\begin{figure}
	\centering
	\includegraphics[width=\columnwidth,trim=0.0cm 0.0cm 0.0cm 0.0cm, clip=true]{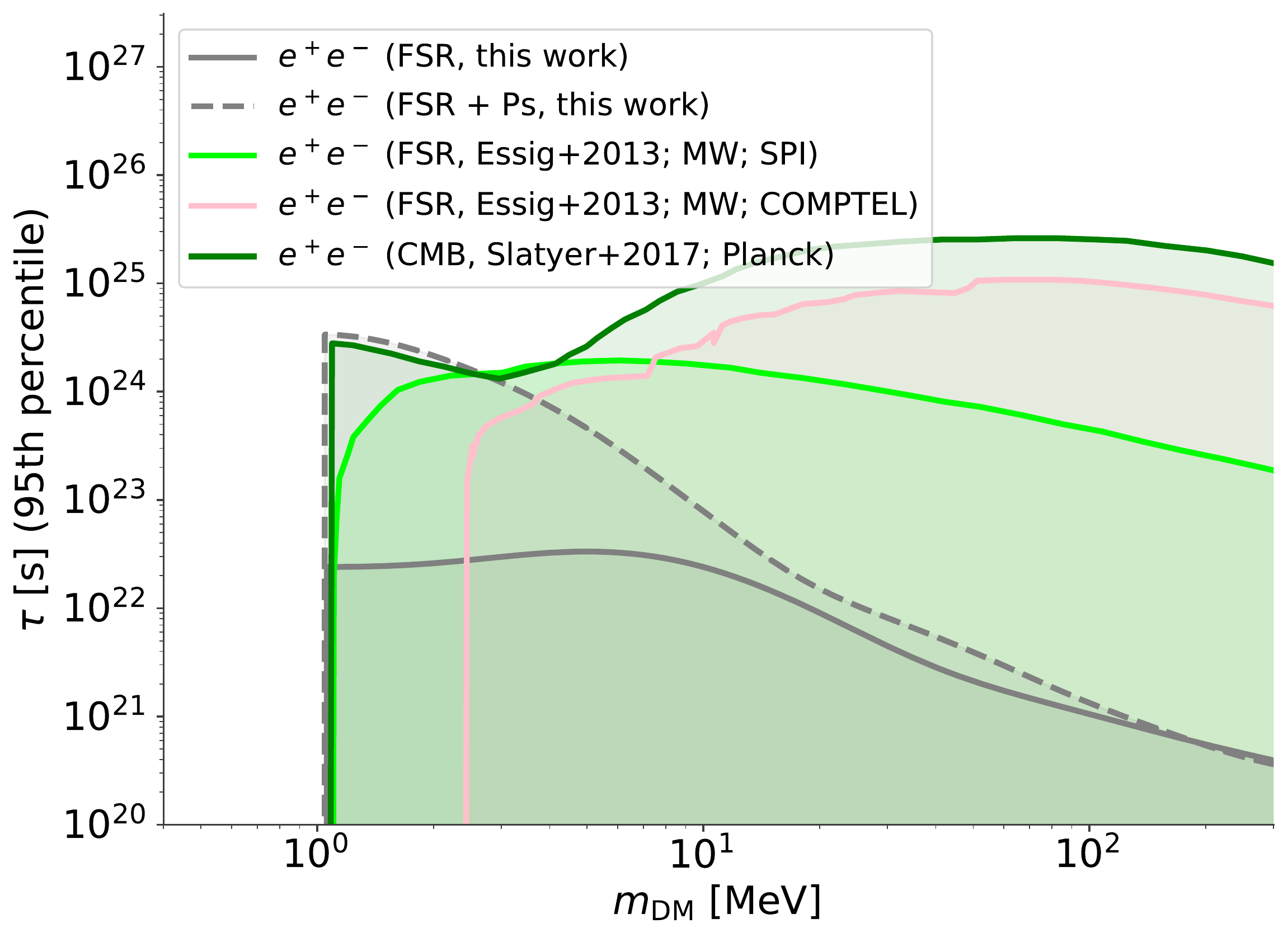}%
	\caption{Similar to Fig.\,\ref{fig:DMAnn_FSR_limits} but for $\mathrm{DM \rightarrow e^+ + e^- + (\gamma)_{\rm FSR}}$.\label{fig:DMDecay_FSR_limits}}
\end{figure}
\begin{figure}
	\centering
	\includegraphics[width=\columnwidth,trim=0.0cm 0.0cm 0.0cm 0.0cm, clip=true]{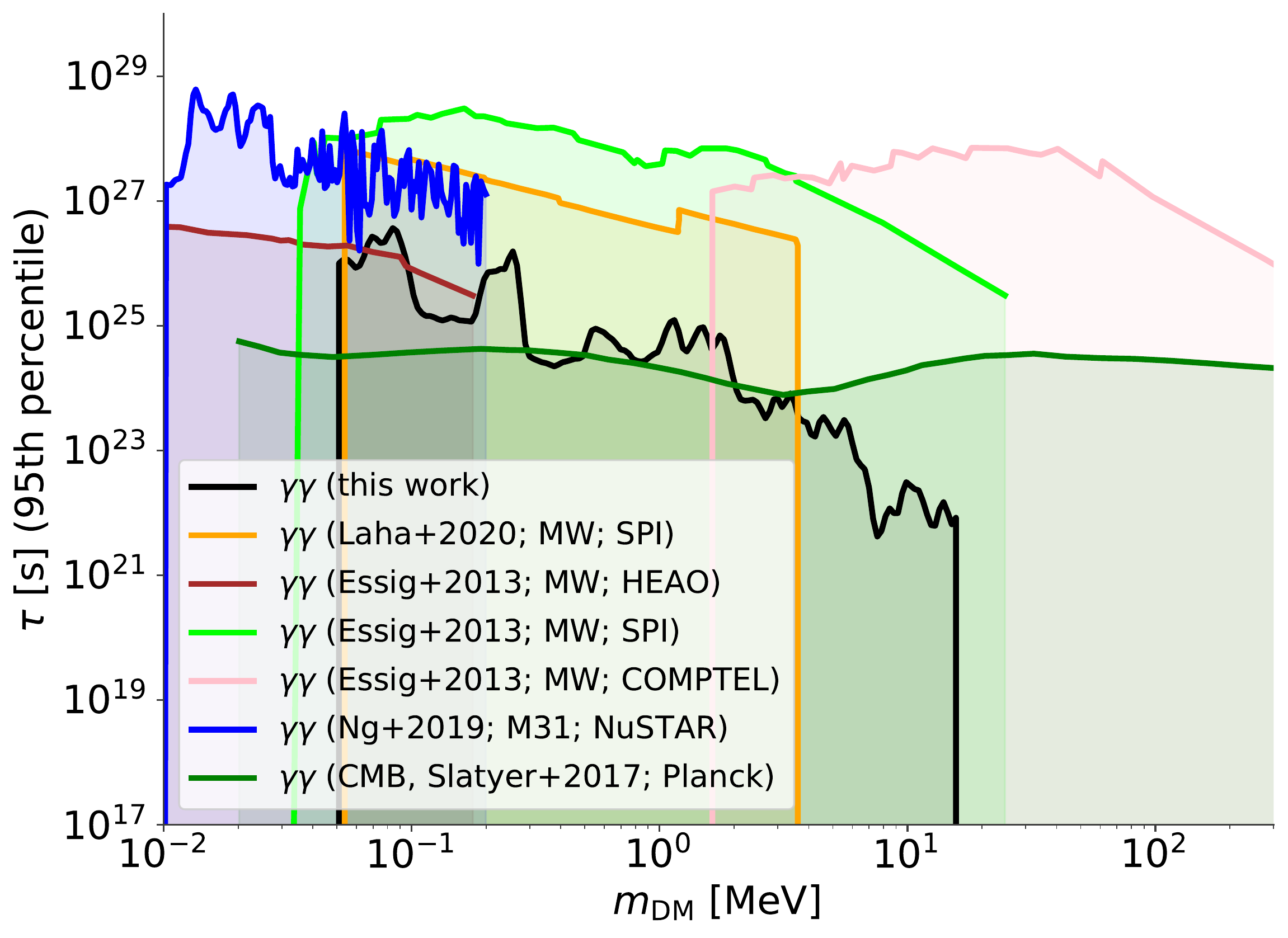}%
	\caption{Similar to Fig.\,\ref{fig:DMAnn_FSR_limits} but for $\mathrm{DM \rightarrow \gamma + \gamma}$. For the case $\mathrm{DM \rightarrow \gamma + Y}$, the limits are smaller by a factor of $2$.\label{fig:DMDecay_gammagamma_limits}}
\end{figure}

In Fig.\,\ref{fig:DMDecay_FSR_limits} we show the the exclusion region of $\tau$ as a function of the DM mass.
Similar to the DM annihilation case, FSR alone is barely constraining, but can be pushed by including $e^+$ annihilation.
We find that DM particles decaying into $e^+e^-$ pairs must have a lifetime longer than $1$--$5 \times 10^{24}$\,s between $1$ and $2\,\mathrm{MeV}$.
In the mass range $1$--$2\,\mathrm{MeV}$, our bounds are as strong as the limits from the CMB \cite{Slatyer2017_CMB_DMdecay}.
Above this mass range, previous limits from SPI and COMPTEL in the MW \citep{Essig2013_DMlimits_gammarays}, and in particular those from the CMB are more constraining.

Finally, our DM decay time limits considering the direct decay into two photons is not improving beyond previous constraints.
This is expected from the pure narrow line sensitivity of SPI.

\section{Summary and Conclusion}\label{sec:conclusion}
Ret\,II is not seen in soft $\gamma$-rays between 0.025 and 8\,MeV, up to an upper flux limit of $10^{-8}\,\mathrm{erg\,cm^{-2}\,s^{-1}}$ ($5 \times 10^{-10}\,\mathrm{erg\,cm^{-2}\,s^{-1}}$ within 20--80\,keV).
The previously reported 511\,keV line signal in Ret\,II has not been consolidated in this work, and we provide an upper limit on its flux of $1.7 \times 10^{-4}\,\mathrm{ph\,cm^{-2}\,s^{-1}}$.
Given the distance to the galaxy, this constrains the $e^+$-annihilation rate (including Ps formation and annihilation in flight) to $0.9$--$3.7 \times 10^{43}\,\mathrm{e^+\,s^{-1}}$, at most the annihilation rate of the MW\footnote{We note that in \citet{Siegert2016_dsph}, there is a typo, reducing the estimate for Ret\,II by a factor of ten -- the actual annihilation rate in the previous work should be on the order of $10^{44}\,\mathrm{e^+\,s^{-1}}$.}.

We calculated parametrised spectral models for PBH evaporation and DM annihilation/decay with subsequent $e^+$ annihilation and fitted these models to the total spectrum from Ret\,II.
Our conservative limit on the monochromatic PBH mass distribution constituting the entirety of DM in Ret\,II is $0.8 \times 10^{16}$\,g.
We improve the limits on the velocity-averaged DM annihilation cross section of decay time of DM particles into $e^+e^-$ in the range $0.5$--$2\,\mathrm{MeV}$.
For annihilation or decay into two photons, our limits are weaker than previous estimates.
We provide the currently only limits on the cross section $\langle \sigma v \rangle_{\gamma\gamma}$ in the range $3$--$8\,\mathrm{MeV}$, ranging between $10^{-28}$ and $10^{-25}\,\mathrm{cm^3\,s^{-1}}$.

While our estimates are at most on the same level as previous constraints, we arrive at our conclusions with a very small data set (two weeks) compared to the multi-year or even decade-long observations.
With our detailed spectral modelling including the effects of possible $e^+$ annihilation and the proper statistical treatment of known unknowns, we show that dedicated observations of selected targets provide valuable information about the DM conundrum.
We encourage follow-up observations of Ret\,II and other dwarf galaxies with INTEGRAL, and foresee a bright future for soft $\gamma$-ray instruments to come, such as COSI \citep{Tomsick2019_COSI}, to contribute significantly in the research for the nature of DM.

\section*{Software}
\textit{OSA/spimodfit} \citep{spimodfit},
\textit{BlackHawk} \citep{Arbey2019_BlackHawk},
\textit{numpy} \citep{Oliphant2006_numpy},
\textit{matplotlib} \citep{Hunter2007_matplotlib},
\textit{astropy} \citep{astropy2013_astropy},
\textit{scipy} \citep{Virtanen2019_scipy},
\textit{3ML} \citep{Vianello2015_3ML},
\textit{MultiNest} \citep{Feroz2008_multinest,Feroz2009_multinest,Feroz2019_multinest}.

\section*{Acknowledgements}
Thomas Siegert is supported by the German Research Foundation (DFG-Forschungsstipendium SI 2502/3-1).
We thank Ranjan Laha, Kenny Ng, Shunsaku Horiuchi, Marco Cirelli, Sam McDermott, Tracy Slatyer, and Joaquim Iguaz for providing limits on dark matter properties, and Oleg Korobkin for $\gamma$-ray spectra from neutron star mergers.
We thank John Beacom and Hassan Y\"uksel for details on the in-flight annihilation spectrum.

\section*{Data Availability}
The data underlying this article will be shared on reasonable request to the corresponding author.



\bibliographystyle{mnras}
\bibliography{thomas.bib} 




\appendix

\section{Details on Spectral Fits}\label{sec:appendix_spec_fits}
The error bars in the extracted spectrum in Sec.\,\ref{sec:total_spec}, Fig.\,\ref{fig:spectrum} approximately follow a normal distribution.
Therefore, we use the likelihood
\begin{equation}
	\mathscr{L}_{\rm normal}(D|M(\boldsymbol \psi)) = \prod_{i=1}^{30} \frac{1}{\sqrt{2\pi}\sigma_i} \exp\left(-\frac{1}{2} \left[ \frac{y_i-m_i(\boldsymbol{\psi})}{\sigma_i} \right]^2 \right)\mathrm{,}
	\label{eq:normal_likelihood}
\end{equation}
\noindent with $y_i$ as the measured flux in energy bin $i=1 \dots 30$ in the spectrum, $\sigma_i$ the corresponding uncertainties, and $m_i(\boldsymbol{\psi})$ the forward-folded model, Eq.\,(\ref{eq:folded_model}), that depends on a set of spectral parameters $\boldsymbol{\psi}$.

Since we want to include the astrophysical and particle physics uncertainties of our source, we construct the joint posterior of all $N$ parameters $\psi_1 \dots \psi_N$ through Bayes' theorem as
\begin{equation}
	\pi(\psi_1 \dots \psi_N|y_i,\sigma_i) \propto \mathscr{L}_{\rm normal}(y_i,\sigma_i | \psi_1 \dots \psi_N) \pi(\psi_1 \dots \psi_N)\mathrm{.}
	\label{eq:Bayes_theorem}
\end{equation}
Here, $\pi(\psi_1 \dots \psi_N)$ is the joint prior distribution of the individual parameters.
We use independent priors so that $\pi(\psi_1 \dots \psi_N) = \pi(\psi_1) \cdots \pi(\psi_N)$, that is, each parameter obtains its individual prior probability distribution function.
By integrating out the parameters $\psi_3$ to $\psi_N$, for example, we can construct the marginalised joint posterior distribution of $\psi_1$ and $\psi_2$,
\begin{equation}
	\pi(\psi_1, \psi_2 | y_i,\sigma_i) = \int \dots \int\,d\psi_2 \cdots \psi_N \pi(\psi_1 \dots \psi_N|y_i,\sigma_i)\mathrm{.}
	\label{eq:marginalised_posterior}
\end{equation}
To obtain the upper bound on $\psi_1$ as a function of $\psi_2$, we integrate the marginalised posterior $\pi(\psi_1, \psi_2 | y_i,\sigma_i)$ in $\psi_1$-direction and equate to a certain percentile $P$,
\begin{equation}
	P = \int_{\psi_1^{\rm min}}^{\psi_1^{\rm ub}(\psi_2)}\,d\psi_1 \pi(\psi_1, \psi_2 | y_i,\sigma_i)\mathrm{.}
	\label{eq:Bayesian_upper_bound}
\end{equation}
\noindent Solving Eq.\,(\ref{eq:Bayesian_upper_bound}) for $\psi_1^{\rm ub}(\psi_2)$ provides the upper bound on $\psi_1$ as function of $\psi_2$.
\begin{figure}
	\centering
	\includegraphics[width=\columnwidth,trim=0.0cm 0.0cm 0.0cm 0.0cm, clip=true]{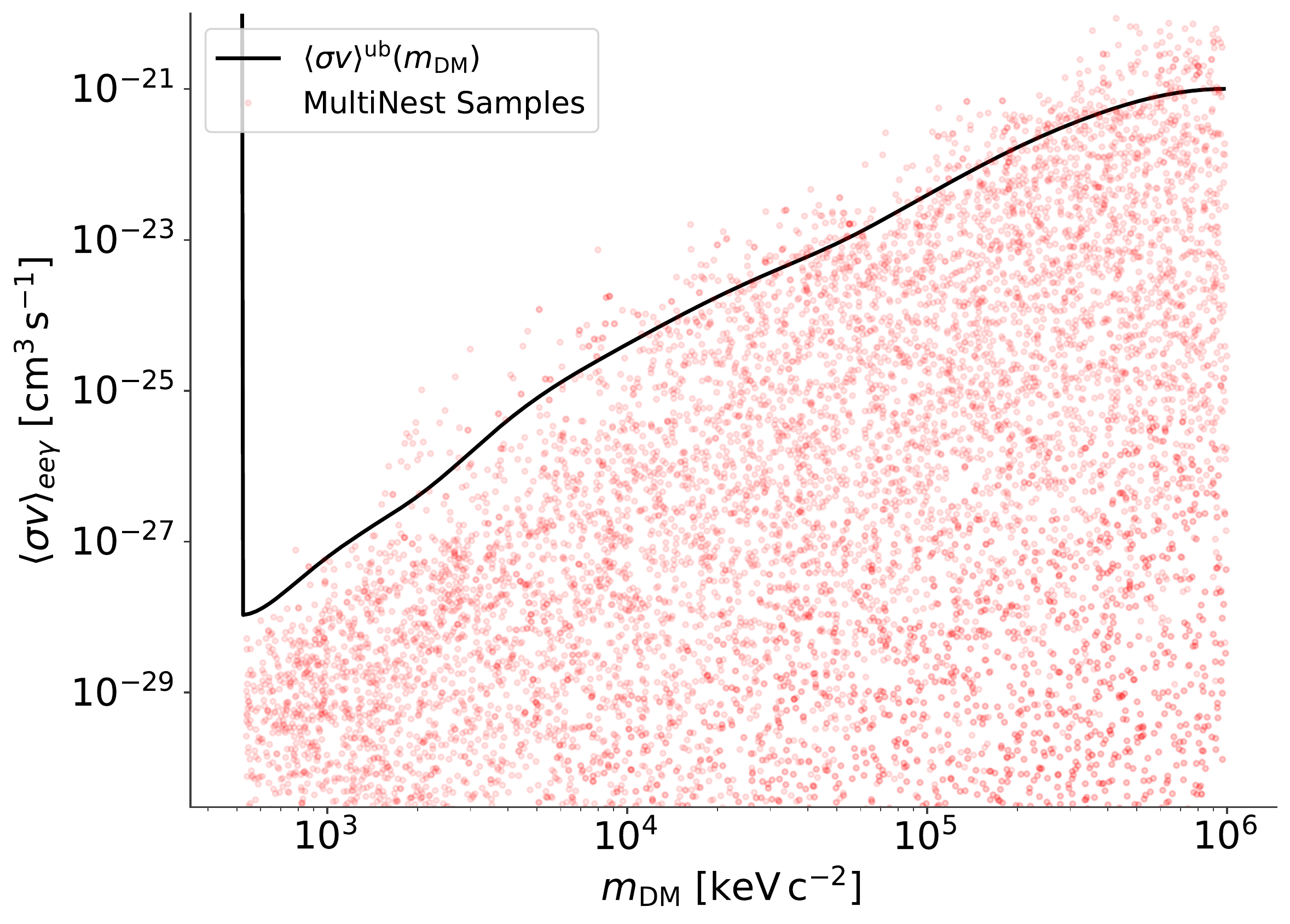}%
	\caption{Marginalised posterior distribution $\pi(\langle \sigma v \rangle, m_{\rm DM}| y_i, \sigma_i)$ (red dots), together with upper bounds of the annihilation cross section as a function of the DM mass (black line).\label{fig:marginalised_posterior_DMAnn}}
\end{figure}

A written example considering the case of annihilating DM into $e^+e^-$ with subsequent FSR and $e^+$ annihilation would calculate the upper bound on the velocity-averaged annihilation cross section $\langle \sigma v \rangle$ as a function of the DM mass $m_{\rm DM}$.
Thus first, the join posterior $\pi(\langle \sigma v \rangle, m_{\rm DM}, J, f_{\rm Ps}, E_{\rm kin}^{\rm max} | y_i, \sigma_i)$ is estimated, then marginalised over the J-factor, the Ps fraction and the kinetic energy threshold to avoid escape,
\begin{eqnarray}
	& \pi(\langle \sigma v \rangle, m_{\rm DM} | y_i, \sigma_i) & = \nonumber\\
	& \int_{J_{\rm min}}^{J_{\rm max}}\,dJ\, \int_{f_{\rm Ps, \rm min}}^{f_{\rm Ps, \rm max}}\,df_{\rm Ps}\, \int_{E_{\rm kin, \rm min}^{\rm max}}^{E_{\rm kin,\rm max}^{\max}}\,dE_{\rm kin}^{\rm max} & \, \nonumber\\
	& \pi(\langle \sigma v \rangle, m_{\rm DM}, J, f_{\rm Ps}, E_{\rm kin}^{\rm max} | y_i, \sigma_i)\mathrm{,} &
	\label{eq:marginalised_posterior_DMAnn}
\end{eqnarray}
\noindent and finally integrated up to the bound $\langle \sigma v \rangle^{\rm ub}(m_{\rm DM})$ by
\begin{equation}
	P = \int_{0}^{\langle \sigma v \rangle^{\rm ub}(m_{\rm DM})}\,d\langle \sigma v \rangle \pi(\langle \sigma v \rangle, m_{\rm DM}  | y_i,\sigma_i)\mathrm{.}
	\label{eq:upper_bound_DMAnn}
\end{equation}
We use \verb|MultiNest| \citep{Feroz2008_multinest,Feroz2009_multinest,Feroz2019_multinest} in 3ML to evaluate the joint posterior distributions, marginalisations, and upper bounds.
We show the samples of this example in Fig.\,\ref{fig:marginalised_posterior_DMAnn}.
We list our prior distributions and parameter ranges for all considered models in Tab.\,\ref{tab:priors}.
\begin{table*}
	\centering
	\resizebox{\textwidth}{!}{
		\begin{tabular}{l|cccccc}
			\hline\hline
			Model & $\kappa$ & $M$ & $D$ & $J$ & $f_{\rm Ps}$ & $E_{\rm kin}^{\rm max}$ \\
			\hline
			PBH (prim.) & $\mathscr{U}(0,1)$ & $\log\mathscr{U}(10^{14},10^{18})$ & $\mathscr{U}(10^{18}, 4 \times 10^{18})$ & -- & -- & --  \\
			PBH (sec.) & \vdots & \vdots & \vdots & -- & -- & --  \\
			PBH (sec.) + Ann. & \vdots & \vdots & \vdots & -- & $\mathscr{U}(0,1)$ & --  \\
			PBH (sec.) + lim. Ann.& \vdots & \vdots & \vdots & -- & \vdots & $\log\mathscr{U}(10^{3},10^{5})$  \\
			\hline
			$\mathrm{2DM \rightarrow e^+e^-\gamma}$ & $\log\mathscr{U}(3 \times 10^{-34}, 3 \times 10^{-20})$ & $\log\mathscr{U}(511,10^{6})$ & -- & $\mathscr{U}(2 \times 10^{18}, 4 \times 10^{19})$ & $\mathscr{U}(0,1)$ & $\log\mathscr{U}(10^{3},10^{5})$ \\
			$\mathrm{2DM \rightarrow 2\gamma}$ & \vdots & $\log\mathscr{U}(25,8000)$ & -- & \vdots & -- & -- \\
			$\mathrm{DM \rightarrow e^+e^-\gamma}$ & $\log\mathscr{U}(10^{15}, 10^{24})$ & $\log\mathscr{U}(1022,10^{6})$ & $\mathscr{U}(10^{18}, 4 \times 10^{18})$ & -- & $\mathscr{U}(0,1)$ & $\log\mathscr{U}(10^{3},10^{5})$ \\
			$\mathrm{DM \rightarrow 2\gamma}$ & $\log\mathscr{U}(10^{15}, 10^{29})$ & $\log\mathscr{U}(50,16000)$ & \vdots & -- & -- & -- \\
			\hline
	\end{tabular}}
	\caption{Prior distributions for considered DM models in Eq.\,(\ref{eq:astroflux}).}
	\label{tab:priors}
\end{table*}


\bsp	
\label{lastpage}
\end{document}